\documentclass[conference,compsoc]{IEEEtran}
\IEEEoverridecommandlockouts
\usepackage{cite}
\usepackage{amsmath,amssymb,amsfonts}
\usepackage{algorithmic}
\usepackage{graphicx}
\usepackage{textcomp}
\usepackage{xcolor}
\usepackage{tikz}
\usepackage{listings}
\usepackage{stfloats}
\usepackage{multirow}
\usepackage{booktabs}
\usepackage{xspace}
\usepackage{xcolor,colortbl}
\usepackage{pifont}
\usepackage{caption}
\usepackage[pdfborder={0 0 0}, citecolor=blue, linkcolor=blue, urlcolor=black, colorlinks=true]{hyperref}
\usepackage[square,numbers,compress]{natbib}
\usepackage[ruled,linesnumbered]{algorithm2e}

\newcommand{\tab}{\hspace*{1em}}

\newcommand{\code}[1]{{\fontfamily{cmtt}\fontseries{m}\fontshape{n}\selectfont\small{#1}}}
\newcommand{\sysname}{\code{Kangaroo}\xspace}
\newcommand{\sysnamenospace}{\code{Kangaroo}}
\newcommand{\squirrel}{{Squirrel}\xspace}
\newcommand{\sqlright}{{SQLRight}\xspace}
\newcommand{\sqlancer}{{SQLancer}\xspace}

\newcommand{\squirrelnospace}{{Squirrel}}
\newcommand{\sqlrightnospace}{{SQLRight}}
\newcommand{\sqlancernospace}{{SQLancer}}

\begin{document}

\definecolor{codegreen}{rgb}{0,0.6,0}
\definecolor{mygray}{gray}{0.9}

\newsavebox\CBox
\def\textBF#1{\sbox\CBox{#1}\resizebox{\wd\CBox}{\ht\CBox}{\textbf{#1}}}

\title{Detecting DBMS Bugs with Context-Sensitive Instantiation and Multi-Plan Execution}

% \and
% \and
% \and
% \and
% \and

\maketitle

\begin{abstract}
DBMS bugs can cause serious consequences, posing severe security and privacy concerns.
This paper works towards the detection of memory bugs and logic bugs in DBMSs, and aims to solve the two innate challenges, including how to generate semantically correct SQL queries in a test case, and how to propose effective oracles to capture logic bugs. 
To this end, our system proposes two key techniques. The first key technique is called context-sensitive instantiation, which considers all static semantic requirements (including but not
limited to the identifier type used by existing systems) to generate semantically valid SQL queries. The second key technique is called multi-plan execution, which can effectively capture logic bugs. Given a test case, multi-plan execution makes the DBMS execute all query plans instead of the default optimal one, and compares the results. A logic bug is detected if a difference is found among the execution results of the executed query plans.
We have implemented a prototype system called \sysname and applied it to three widely used and well-tested DBMSs, including SQLite, PostgreSQL, and MySQL.
Our system successfully detected 50 new bugs. The comparison between our system with the state-of-the-art systems shows that our system outperforms them in terms of the number of generated semantically valid SQL queries, the explored code paths during testing, and the detected bugs.
\end{abstract}

\begin{IEEEkeywords}
DBMS, bug finding, Fuzzing
\end{IEEEkeywords}

%-------------------------------------------------------------------------------
\section{Introduction}
%-------------------------------------------------------------------------------

% Database management system (DBMS), as a fundamental infrastructure, has been widely used in various systems~\cite{sqliteusers, mysqlusers}.
Database management systems (DBMSs) provide fundamental infrastructure for many applications~\cite{sqliteusers, mysqlusers}, so it is crucial that they can be relied upon.
DBMS bugs could result in data leakage, data manipulation, or service termination, and even pose severe security threats~\cite{BUGSQLiteUAF, BUGGoogleChrome, jung2019apollo, zhong2020squirrel}.
Thus, timely detection of DBMS bugs is an emerging need.

%a constant yet essential activity during the entire life cycle of DBMS.
DBMS bugs can be roughly categorized into two types, i.e., memory bugs and logic bugs. 
A memory bug occurs when the DBMS abnormally terminates due to a memory crash or an assertion failure.
% A logic bug occurs when DBMS produces an incorrect output for a given test case~\footnote{In this paper, a test case means a bundle of SQL statements (or SQL queries). SQL statements and SQL queries are used interchangeably.}.
A logic bug occurs when DBMS produces an incorrect output for a given test case.
Recent works show that coverage-guided mutation-based DBMS fuzzing systems~\cite{sqlsmith, chen2021one, zhong2020squirrel, liu2021automated, zhang2021duplicate, ghit2020sparkfuzz} have proven more effective than generation-based DBMS fuzzing systems because they can generate more diverse test cases (as inputs to a DBMS system) to trigger the bugs. 

However, effectively detecting DBMS bugs has two innate challenges.
\textit{The first challenge is how to generate syntactically and semantically correct SQL statements in a test case during the mutation.}
DBMS performs syntactic and semantic checking on the SQL statements to ensure their validity.
An invalid mutated SQL statement will be discarded before being fed into the DBMS execution engine. Thus, the bugs in the execution engine will not be triggered.
\textit{The second challenge is how to propose effective oracles to capture logic bugs that do not cause the program termination}.
Unlike memory bugs that usually cause a crash of the target program (easily captured by memory sanitizers such as ASan~\cite{ASan}), a logic bug does not terminate the program but generates an incorrect result. 

\smallskip
\noindent \textbf{Limitations of Existing Systems}\tab
Existing systems have limitations in addressing these challenges.
First, during the mutation of a test case, they only consider limited semantic constraints.
As a result, they will generate invalid SQL statements that cannot pass the semantic check of a DBMS.
For instance, \squirrel~\cite{zhong2020squirrel} proposed \textit{semantics-guided instantiation} and the follow-up work \sqlright~\cite{liang2022detecting} proposed \textit{Context-based IR instantiation} to improve the semantic correctness of generated SQL statements.
For simplicity, we will refer to their approaches as \textit{type-sensitive instantiation} in this paper since they utilize a context-free strategy that only considers the correctness of the identifier type (e.g., table, column).
% For example, \code{SELECT score FROM (SELECT name FROM student) AS namelist ;} is invalid, even though all identifier types are correct.
% To guarantee the valid rate of mutated SQL statements, existing systems have to limit the complexity of generated queries to tolerate their incomplete constraints.

% \begin{figure}[t]
% \centerline{\includegraphics[width=0.25\textwidth]{Fig/SQLdig-inputspace.png}}
% \caption{The input space that different oracles can explore. PQS, NoREC, and TLP can only explore limited spaces because they have strict requirements for SQL queries. The multi-plan execution has no requirement and thus can explore a wider input space.}
% \label{fig:comparsion-oracle}
% % \vspace{-0.3in}
% \end{figure}

Second, oracles used by existing systems~\cite{rigger2020testing, rigger2020detecting, rigger2020finding, liang2022detecting,tang2023detecting} to detect logic bugs have strict requirements on the SQL statements.
This makes the fuzzing system only explore a narrow space of inputs, leading to limited bugs that can be detected.
For example, 
NoREC~\cite{rigger2020detecting} requires that the effective SQL query in a test case has \code{WHERE} clauses, thus it can only detect logic bugs due to the optimization of \code{WHERE} clauses.
Another recent work, TQS~\cite{tang2023detecting}, focusing on detecting logic bugs caused by equal-join optimization, fails to identify bugs arising from other types of optimization.
Furthermore, both NoREC and TQS attempt to explore multiple query plans of queries but are only capable of covering partial query plans.
Hence, an oracle that can be applied to SQL statements without strict requirements and explore more query plans is needed.

\smallskip
\noindent\textbf{Our Solution}\tab
This work proposes two key techniques to address the limitations and solve the two innate challenges. 
The first key technique is called \textit{context-sensitive instantiation}.
It performs context-sensitive analysis to collect all static semantic constraints (including but not limited to the identifier type) to improve the semantic correctness of generated inputs.
% Specifically, it first translates the parse tree of a mutated SQL query to a semantic tree.
% Each node of the semantic tree represents a SQL clause that is a meaningful logical chunk of the SQL statement.
% Next, it collects semantic constraints by performing context-sensitive analysis on each semantic node.
% After collecting all semantic constraints, it utilizes a random heuristic algorithm to
% solve them and finds a probable solution for all the variables.
% By doing so, our system obtains more accurate and richer semantic constraints.
For instance, a SQL statement \code{SELECT x1+i1 FROM x2 WHERE x3=x4} has four identifiers \code{x1$\sim$x4} and a constant \code{i1}.
\squirrel and \sqlright can only infer that \code{x1, x3} and \code{x4}
are columns of the table \code{x2}. 
In contrast, our system can additionally infer that the type of column \code{x1} must be numeric and the type of columns \code{x3} and \code{x4} should be comparable (e.g., both are string).
With more inferred semantic constraints, our system can generate more valid SQL queries.
We evaluate the effectiveness of our method and found it achieves 1.14x${\sim}$2.87x higher semantic correctness than previous approaches (Section~\ref{sec:query_validity_test}).

The second key technique is called \textit{multi-plan execution} (abbreviated as MPE) which can be applied to different types of SQL queries to detect logic bugs in the DBMS execution engine.
The high-level idea is to compare the results of different query plans of the same SQL query.
Specifically, when a SQL query is submitted to a DBMS, the query optimizer will evaluate various query plans and determine the optimal one to execute.
Our system hooks the process of query optimization to make the DBMS engine execute all the query plans (instead of executing the optimal one in the vanilla DBMS) and compares the results. A logic bug is detected if there is a difference between the results of different query plans. Note that MPE does not put strict limitations on the SQL queries that can be tested, which will improve the explored edges and the number of detected bugs (Section~\ref{sec:unit_test}).

\smallskip
\noindent\textbf{Prototype and Evaluation}\tab
In this study, we implemented a prototype named \sysname and applied it to three widely-used database management systems (DBMSs): SQLite~\cite{sqlite}, PostgreSQL~\cite{postgres}, and MySQL~\cite{mysql}, to evaluate its effectiveness. Despite the fact that state-of-the-art tools have already extensively tested these DBMSs, Kangaroo successfully identified 50 new bugs, consisting of 9 logic bugs, 19 crash-causing bugs, and 22 assertion failure-inducing bugs. As of the time of writing, 28 of these bugs have been fixed, with 11 assigned CVEs~\cite{CVE}. These results demonstrate the efficacy of our system.

We also compared Kangaroo with leading DBMS testing tools, such as \squirrel, \sqlancer, and \sqlright. Our evaluation reveals that \sysname surpasses these tools in terms of generating semantically valid SQL queries, exploring code paths, and detecting bugs. Specifically, after conducting a 24-hour test on the three DBMSs, \sysname successfully detected 17 bugs, while \sqlancer, \squirrel, and \sqlright only identified 1, 3, and 6 bugs, respectively. In addition, \sysname proved more effective in generating valid SQL queries, achieving a 1.6x-1.9x improvement compared to the mutation-based tool \squirrel, and it explored 1.16x-1.3x more program states than the rule-based tool \sqlancer.

% We have implemented a prototype called \sysname and applied
% it to three popular DBMSs, i.e., SQLite~\cite{sqlite}, PostgreSQL~\cite{postgres}, and MySQL~\cite{mysql}, to evaluate its effectiveness. 
% Although state-of-the-art tools have thoroughly tested these DBMSs, we successfully found {50} new bugs, including nine logic bugs, 19 bugs that cause DBMS crashes, and 22 bugs that cause assertion failures.
% As of this writing, 28 have been fixed with 11 assigned CVEs. This result demonstrated the effectiveness of our system.
% We also compared our system with state-of-the-art DBMS testing tools, including \squirrel, \sqlancer, and \sqlright. 
% Our evaluation shows that \sysname outperforms them regarding the number of generated semantically valid SQL queries, the explored code paths, and the detected bugs.
% % in terms of the explored execution paths and detected bugs. 
% Specifically, 
% after running a 24-hour test on three DBMSs, \sysname successfully detected 17 bugs, while \sqlancer, \squirrel, and \sqlright found only 1, 3, and 6 bugs, respectively. 
% % It is worth noting that
% Besides, \sysname is more effective in generating valid SQL queries (1.6x-1.9x improvement) than the mutation-based tool \squirrel, and it can explore more program states (1.16x-1.3x paths) than the rule-based tool \sqlancer.

This work makes the following main contributions. 
\begin{itemize}

    \item We revealed the challenges of effectively detecting DBMS bugs and the limitations of existing solutions.

    \item We proposed two key techniques to solve the challenges, including context-sensitive instantiation to improve the semantic correctness of the mutated SQL queries and the MPE that can be applied to different types of SQL queries to detect logic bugs.
     % \item We solved three technical challenges, including the way to semi-automatically generate parsers, the semantic validation to improve the semantic validity of mutated SQL queries, and a new sanitizer called QueryPlan Sanitizer. 
   
    \item We implemented and applied a prototype system
          to three popular DBMSs. Our system successfully detected 50 new bugs. The micro-benchmark also shows our system outperforms existing ones in generating semantically valid SQL queries.
          
\end{itemize}

We plan to release the code of \sysname to help DBMS developers enhance the security of their products.
% We plan to release the code of \sysname \footnote{Link of \sysname prototype:~\url{https://github.com/anonymous44117/Kangaroo}} to help DBMS developers enhance the security of their products.

%-----------------------
\section{Background}
%-----------------------

\subsection{Structured Query Language}\label{sec:sql}
Structured Query Language (SQL) is a domain-specific language used to manage data held in DBMS. 
SQL statements are the smallest execution unit that typically consists of one or several SQL clauses, e.g., \code{WHERE} clause, \code{FROM} clause, \code{ORDER BY} clause, etc.
Each SQL clause is a meaningful logical chunk that consists of identifiers, constants, SQL keywords, and sub-clauses.
% In other words, the smallest units in SQL statements are identifiers, constants, and SQL keywords.
\textit{We refer identifiers and constants to variables in this paper} because their values are changeable, and these changes do not affect the syntactic structure.
The semantics of variables are determined by the type of clauses they directly or indirectly belong to and the patterns of the clauses. 
In this paper, we denote these semantics as context information.
Since our semantic analysis considers the context information, we call it \textit{context-sensitive analysis}.

\subsection{SQL Query Processing}
Most relational DBMSs use Structured Query Language (SQL)~\cite{chamberlin1974sequel} for querying and maintaining the database.
Any SQL queries will be executed in the following steps.

{First}, {a DBMS parser} performs lexical analysis to divide the query
	into meaningful chunks, called tokens. Then, it does syntax analysis to figure out
	the syntax structure of these tokens and generate a raw parser tree.
	It checks for syntax validity and terminates execution if
	any syntax errors are detected.
After that, {the translator} of DBMS analyzes the raw parser tree, checks its semantics,
	and converts it into an internal representation that can be used by
	the {optimizer} and the {executor}. 
Then
%Since database structures are complex, in most cases, the needed data for a statement can be collected from a database by accessing it in different ways through different data structures, and in different orders.  
	the optimizer (also called planner) will try to find all possible query plans for a SQL statement and select the optimal one by evaluating the cost of each query plan.
At last, the executor of DBMS
	runs the optimal query plan and returns the required result. 
The executor also checks whether the dynamic semantics are correct during execution. 
 % For example, the executor will check whether a scalar subquery actually returns a single value after the subquery is executed.

In conclusion, DBMS can be divided into four main components, i.e., parser, translator, planner, and executor.
The planner and executor, as the core parts of a DBMS, are the most complex components.
Thus, it is crucial to make the test cases reachable to these two modules.

\subsection{Levels of DBMS Testing}

\begin{table}[t]
\caption{Eight levels of SQL statement correctness and the corresponding reachable DBMS modules.
% The last column shows the theoretical lower bound of the correctness level for different systems' generated input. Our system can ensure the generated input is static confirming with the help of context-sensitive instantiation (Section~\ref{subsec:context_sensitive_instantiation}).
} % title of Table
\scriptsize
\centering % used for centering table
\setlength{\tabcolsep}{0.85mm}{
\begin{tabular}{c l l l l} % centered columns (4 columns)
% \hline %inserts double horizontal lines  % inserts table
\toprule
Level & Input class & Module & Incorrect input examples & Fuzzer \\
%heading
\hline
1 & Binary input & Lexer & - & AFL \\
2 & Sequence of ASCII & Lexer & Binary input & \\
3 & Sequence of words &  Parser & Incorrect SQL keywords & \\
4 & Syntactically correct & Translator & Missing semicolon & \\
5 & Identifier type correct & Translator & Query a trigger & Squirrel \\  
	&  &  &  & SQLRight \\  
6 & Data type correct & Translator & Add an integer to a string& \\
7 & Statically confirming & Executor & Ambiguous column names & \textbf{Our system} \\
8 & Dynamically confirming & Executor & Scalar query returns & \\
	& (Semantic correctness)  & & multiple values & \\
% \hline %inserts single line
\bottomrule
\end{tabular}}
\label{tab:testcase_levels}
% \vspace{-0.3cm}
\end{table}
	
The correctness of a SQL statement can be divided into eight levels as shown in Table~\ref{tab:testcase_levels}.
For a syntactically correct SQL statement, the value of variables in the statement determines whether the statement is semantically correct.
The statement's context information imposes various restrictions, including \textit{static constraints} checked by the translator and \textit{dynamic constraints} checked by the executor.
For example, in the statement \code{SELECT 1==(SELECT a FROM t1)}, \code{a} should be numeric, which is a static constraint. Besides, the subquery \code{SELECT a FROM t1} should return no more than one row, which is a dynamic constraint.
A statement is semantically correct (level-8) if all constraints are satisfied.
If all static constraints are satisfied but some dynamic constraints are not, the statement is statically confirming (level-7).
Static constraints can be further divided into different levels.
Identifier type correct (level-5) indicates that all identifiers in SQL statements have correct types (e.g. Column) and the data type correct (level-6) guarantees the data type (e.g. integer) of variables is correct.

Different systems can guarantee different correctness levels.
For example, general-purpose fuzzers such as AFL~\cite{AFL} are inefficient for DBMS testing 
since they lack more sophisticated approaches to generate higher-level correct inputs.
Benefiting from syntax-preserving mutations and naive constraint-guided instantiation, \squirrel is more likely to generate higher-level correct input, thus significantly improving the ability to detect bugs.
% However, when the test cases are complex or the DBMS semantic checks are strict, \squirrel does not perform well because of its incomplete and inaccurate constraints.
% However, \squirrel has to limit the complexity of generated queries to tolerate its incomplete and inaccurate constraints because the complexity of constraints dramatically increases with the complexity of the statement and the strictness of the DBMS checks.
However, \squirrel has to limit the complexity of generated queries to tolerate its incomplete and inaccurate constraints.
What's worse, the complexity of constraints dramatically increases with the complexity of the statement and the strictness of the DBMS checks.
To better explore the core modules of DBMS,
we propose a new method to generate more diverse and complex test cases,
while at the same time, they can be statically confirming (level-7).

\subsection{DBMS Testing}\label{DBMS_Fuzzing}
% 基于生成的Fuzzing策略

\smallskip
\noindent\textbf{Test Case Generation}\tab
DBMS testing aims to trigger bugs by constructing abundant test cases.
There are two methods to generate SQL queries. 
The rule-based ones~\cite{sutton2007fuzzing,tang2023detecting,rigger2020detecting,rigger2020finding,rigger2020testing} generate test cases following a predefined model to ensure the generated SQL queries can pass the SQL parser.
However, building a precise model requires domain knowledge.
Besides, the generated inputs cannot efficiently explore
the program's state space since it wastes much effort on similar queries.

% Compared with the previous one,
The mutation-based method~\cite{neystadt2008automated,zhong2020squirrel,liang2022detecting} generates new test cases by
mutating seed queries.
However, the general mutation strategies, like flipping bits,
can not generate valid SQL queries that can pass the SQL parser.
To address this issue, a common approach is to perform mutations
based on grammar rules~\cite{zhong2020squirrel, wang2019superion}.
It first generates a syntax tree for a given input, then creates specific mutations
by using mutation operators on a tree node.
Although the grammar-based mutation can always generate valid test cases,
 this approach requires non-negligible expert knowledge and manual efforts to
 implement an accurate parser for the target DBMS.

\smallskip
\noindent\textbf{Oracle}\tab
An oracle is a mechanism for determining whether the actual outputs match the expected outcomes.
Differential testing uses different implementations of the same functionality as cross-referencing oracles.
It provides the same inputs to a series of similar systems and then obverses the results.
Any inconsistency between the results may indicate a potential bug.
RAGS~\cite{slutz1998massive} is the first work that applies
differential testing to find logic bugs for DBMS.

Metamorphic testing addresses the test oracle problem based on the observation that a transformation of the input has a known effect on the output.
NoREC~\cite{rigger2020detecting} is an oracle for logic bug detection.
It translates a query that is potentially optimized by DBMS to a query that can hardly be optimized.
Although this approach has been effective in detecting bugs
in widely-used DBMS~\cite{manuelriggerbugs}, it can only be applied to a subset of SQL that can be translated.
% That is to say, the input space of the SQL queries that can be tested is narrow.

Our system leverages the idea of differential testing to \textit{compare the execution results of multiple query plans}, hence named \textit{multi-plan execution} (MPE).

% \smallskip
% \noindent\textbf{Security Impact}\tab

%-------------------------------------------------------------------------------
\section{Motivating Examples}
%-------------------------------------------------------------------------------

In this section, we use two real examples to demonstrate the advantages of our system's two key techniques.

\subsection{Context-Sensitive Instantiation}

\begin{figure*}[t]
\centerline{\includegraphics[width=1.0\textwidth]{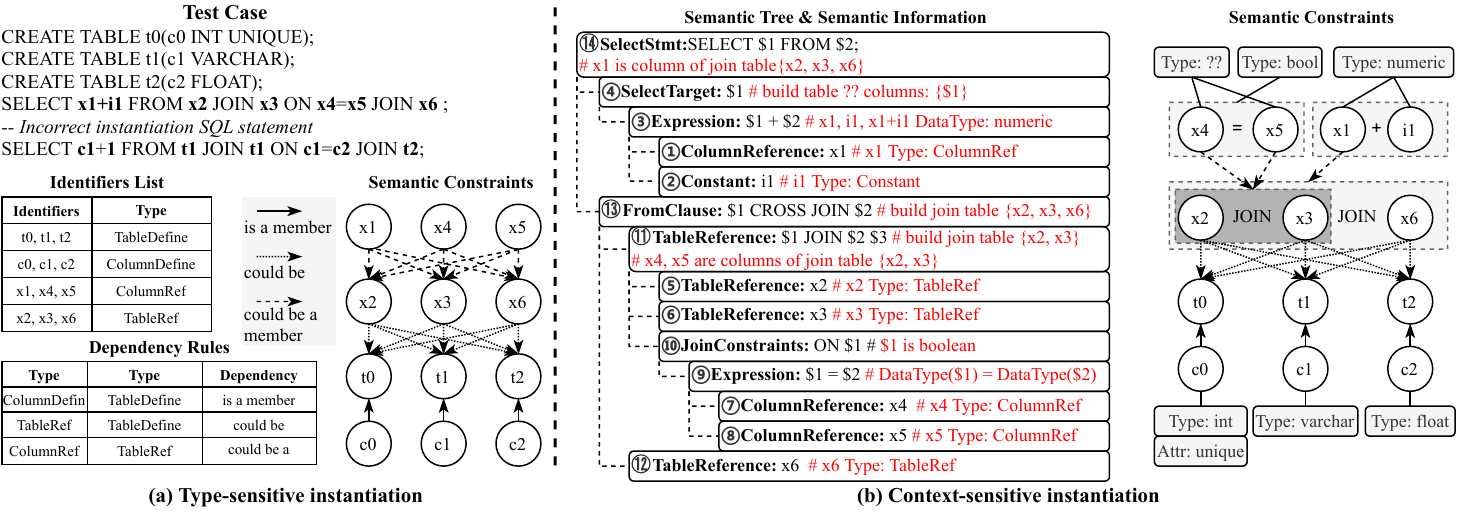}}
\caption{The comparison  between \textit{type-sensitive instantiation} used by existing systems and \textit{context-sensitive instantiation} used in our system. Figure(a) shows semantic constraints inferred from identifier types and manually defined rules. Figure(b) illustrates the SQL clause structure (i.e., semantic tree) of the SELECT statement and the semantic constraints we collect in each clause. In each SQL clause, the child clauses are numbered \$1, \$2, and so forth. The content after '\#' indicates the semantic information obtained from that clause.}
\label{fig:comparsion-instantiation}
% \vspace{-0.3cm}
\end{figure*}

We propose context-sensitive instantiation because of two observations. First, previous works~\cite{zhong2020squirrel, liang2022detecting} have proven the importance of semantic correctness to fuzzing efficiency, but existing type-sensitive instantiation still has large room for improvement.
Second, MPE requires statements to be semantically correct. Our experiments (Section ~\ref{sec:unit_test}) found that the relatively low semantic correctness of existing type-sensitive instantiation cannot fully take advantage of MPE.

\smallskip
\noindent\textbf{Existing Solution}\tab
% Figure~\ref{fig:comparsion-instantiation} shows a test case consisting of multiple SQL queries.
Type-sensitive instantiation uses context-free analysis to get the types of identifiers, then infers their dependencies according to manually predefined rules.
However, these semantic constraints are incomplete and inaccurate, thus insufficient to generate valid SQL queries.
First, the dependencies between identifiers are not only associated with their types but also depend on the context information.
Second, the semantic requirements for a symbolized statement are not limited to dependencies between identifiers.
These identifiers may have various attributes that need to be satisfied, and the constants in statements may also have value requirements.

For example, Figure~\ref{fig:comparsion-instantiation}a shows the dependencies of the example test case inferred from identifier types and manually defined rules.
% the type of all identifiers in the example test case and the inferred dependencies between these identifiers.
The concrete statement \code{SELECT c1+1 FROM t1 JOIN t1 ON c1=c2 CROSS JOIN t2;} suffers from three different semantic errors even though it satisfies all the dependencies considered by type-sensitive instantiation.
First, it is illegal to concrete identifier \code{x5} with column \code{c2}.
Because \code{x5} under the context of the \code{join constraints} of the joined table \code{"x2 JOIN x3"}, \code{x5} could only be the column of table \code{x2} or \code{x3}.
The neglect of this context information leads to a semantic error.
Second, when \code{x2} and \code{x3} are concreted with the same table \code{t1}, there are two columns with the same name in the joined table. Thus, concreting \code{x4} with \code{c1} will cause ambiguous column names error.
Besides, the statement adds the column \code{c1}, which is a string, to an integer.
The absence of the unique name and data type requirements causes these two semantic errors.

\smallskip
\noindent\textBF{Our Solution}\tab
The main difference between context-sensitive instantiation and type-sensitive instantiation is the way semantic constraints are collected.
Context-sensitive instantiation traverses and analyzes all SQL clauses to get complete and accurate static constraints, thus can theoretically generate statically confirmed SQL statements.

Figure~\ref{fig:comparsion-instantiation}b illustrates the structure of all the clauses in the \code{SELECT} statement, the semantic information on each clause, and the inferred semantic constraints.
Compared with type-sensitive instantiation, we can infer the following extra semantic constraints.
In \code{Expression x1+i1}, the data type of column \code{x1} and constant \code{i1} should be numerical since they are the operands of the addition expression.
In \code{Expression x4=x5}, the data type of columns \code{x4} and \code{x5} should be the same type (e.g., both are numeric).
In \code{TableReference x2 JOIN x3}, columns \code{x4} and \code{x5}, which belong to the \code{join} constraint of the joined table, can only be the column of the joined table.
Besides, each column should have a unique name within its dependent table.
These additional constraints guarantee the semantic correctness of this \code{SELECT} statement that does not contain dynamic constraints.
Enforcing dynamic constraints requires emulating the execution of statements, which is challenging.
Thus, our semantic analysis ignores dynamic constraints and leaves it as part of future work.  

\begin{figure}[t]
\centerline{\includegraphics[width=0.48\textwidth]{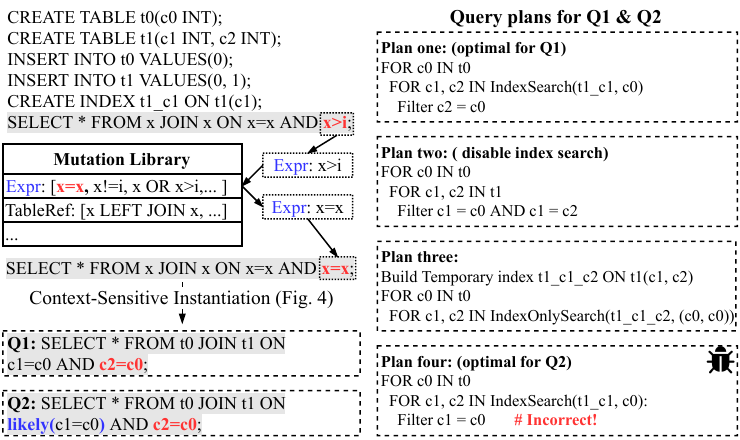}}
\caption{An example that demonstrates the benefits of multi-plan execution. It shows a mutated test case that triggers a logic bug in SQLite. There are four query plans for the mutated SELECT statement Q1. One of them (the fourth one) returns a non-empty result that is different from the others.}
\label{fig:multi-plan-example}
% \vspace{-0.3cm}
\end{figure}

% \begin{lstlisting}[language=sql,float=t,
% label={list:multi-plan-example},
% numbers=left,
% numbersep=2pt,
% xleftmargin=1em, 
% commentstyle=\color{codegreen},
% backgroundcolor=\color{mygray},
% basicstyle=\scriptsize\ttfamily,
% numberstyle=\scriptsize,
% keywordstyle=\bfseries\color{purple},
% % keywordstyle=\bfseries,
% breaklines=true,
% numberstyle=\ttfamily,
% caption={An example that demonstrates the benefits of multi-plan execution. It shows a test case that triggers a logic bug in SQLite. The bug is due to the equivalence transfer optimization. There are four query plans for the SELECT statement. One of them (the fourth one) returns a non-empty result that is different from the others.} ]
% CREATE TABLE t0(c0 INT);
% CREATE TABLE t1(c1 INT, c2 INT);
% INSERT INTO t0 VALUES(0);
% INSERT INTO t1 VALUES(0, 1);
% CREATE INDEX t1_c1 ON t1(c1);
% SELECT * FROM t0 JOIN t1 ON c1=c0 AND c2=c0;
% -- SELECT * FROM t0 JOIN t1 ON likely(c1=c0) AND c2=c0;  -- The slightly different query that can trigger the bug
% /*
% All query plans of buggy query found by MPE
% -- Plan one
% FOR c0 IN t0
%   FOR c1, c2 IN IndexSearch(t1_c1, c0)
%     Filter c2 = c0
% -- Plan two 
% FOR c0 IN t0
%   FOR c1, c2 IN t1
%     Filter c1 = c0 AND c1 = c2
% -- Plan three
% Build Temporary index t1_c1_c2 ON t1(c1, c2)
% FOR c0 IN t0
%   FOR c1, c2 IN IndexOnlySearch(t1_c1_c2, (c0, c0))
% -- Plan four
% FOR c0 IN t0
%   FOR c1, c2 IN IndexSearch(t1_c1, c0):
%     Filter c1 = c0       # Incorrect!
% */
% \end{lstlisting}

% 这一段只点出NoREC和TQS.
\subsection{Multi-Plan Execution}
Although Rigger \textit{et al.} proposed three oracles for DBMS logic bug detection, including 
PQS~\cite{rigger2020testing}, NoREC~\cite{rigger2020detecting}, and TLP~\cite{rigger2020finding}, all of them put limitations on the SQL queries. 
Similarly, TQS~\cite{tang2023detecting}, which aims at detecting logic bugs in equal-join optimization, also limits the diversity of SQL statements. 
% Thus, all of them are unable to comprehensively explore program states and DBMS bugs.  
% As a result, they limit the diversity of SQL statements, which hinders them from comprehensively exploring program states and DBMS bugs. 
In addition, both TQS and NoREC attempt to explore different query plans for queries.
NoREC only compares two distinct plans by transforming queries into semantically similar ones. 
TQS uses DBMS-specific features such as optimization switches and hints to iterate different query plans.
However, this approach is not general and only covers partial query plans.

%since the SELECT statement is one of the most powerful and potentially complex statements within the SQL language.
To this end, we propose MPE, which has no such limitation.
The basic idea is that our system hooks into the DBMS optimizer to execute all query plans and compare their results.
If the result of one query plan is different from the others, a logic bug is detected.

In the following, we use the proof-of-concept of one real-world bug of SQLite in Figure~\ref{fig:multi-plan-example} to illustrate why the existing system cannot explore enough program states to detect the logic bug and how the MPE can capture such a bug.
When the test case is executed, the DBMS optimizer finds four query plans for the \code{SELECT} statement.
Plan one is optimal among four query plans and is executed by SQLite by default.
% In practice, there are typically more than two possible query plans for a query.
%Plan two is optimal when shifting the condition in \code{WHERE} clause to \code{SELECT\_TARGET} clause (line 7).

\smallskip
\noindent\textbf{Why Existing Works Cannot Detect the Bug}\tab
NoREC is one of the most effective DBMS logic bug detection oracles which requires the \code{SELECT} statements to satisfy some predefined rules, e.g., having a \code{WHERE} and \code{FROM} clause.
It shifts the condition in \code{WHERE} clause to the \code{SELECT\_TARGET}, and then executes both to compare the execution result. However, the SELECT statement Q1 in Figure~\ref{fig:multi-plan-example} does not meet the requirement (lacks \code{WHERE} clause), thus missing the bug triggered by the test case.

TQS relies on optimization switches and hints to iterate different query plans, which limits its exploration of query plans.
First, the optimization switches cannot force optimization adoptions nor control the scope of optimizations.
Hints enables more fine-grained control than optimization switches, but is only supported by some DBMSs. 
In addition, optimization switches and hints can typically only manipulate a subset of the optimizations, making it challenging to explore all possible query plans.
Since SQLite does not support hints, TQS can only generate plan two in Figure~\ref{fig:multi-plan-example} by turning off index optimization and therefore misses the bug.

\smallskip
\noindent\textBF{Why Our System Can Capture the Bug}\tab
Our system executes \textit{all the query plans} (including the optimal one and all others) for the given test case. 
Therefore, the bug in plan four will be captured by our system since it outputs a non-empty result while the other plans return an empty result.

One may argue that the bug found in query plan four is meaningless since it will not be executed by default.
In other words, the bug cannot be triggered in practice.
However, even though the buggy query plan is not optimal for the \code{SELECT} statement in the test case, it can be the optimal one for a slightly different SQL statement (Q2). 
Actually, almost all bugs detected by MPE can be triggered in practice. 
We discuss exceptions later in the part of limitations of MPE.

The root cause of this bug is the incorrect equivalence transfer optimization. Because the expression in \code{join constraints} clause is \code{c1=c0 AND c2=c0}, which implies \code{c1=c2}. SQLite uses the index on \code{c1} for the constraint on \code{c2} because of this inference.
Specifically, it uses the values of \code{c0} as the key to search index \code{t1\_c1} to fetch all records that satisfy \code{c1=c0} and then tries to judge \code{c1=c0} on these records again.
However, it should check \code{c1=c2} rather than \code{c1=c0}.
As a result, the expression \code{c1=c0 AND c2=c0} is incorrectly optimized to \code{c1=c0}.

From the security perspective, the incorrect optimization in Figure~\ref{fig:multi-plan-example} will cause the data filter condition to be ignored and return extra data.
Attackers may exploit this bug to steal information if they can trigger the program to conduct similar bugged queries.

\smallskip
\noindent\textbf{Limitations of MPE}\tab
First, MPE supports all SQL grammar instead of a subset but requires the generated queries to have multiple query plans.
Nevertheless, queries with only one query plan (i.e. \code{SELECT 1}) are generally very simple and of low value for finding bugs.
Second, MPE may lead to false negatives due to the same incorrect answers in all query plans.
Such bugs are rarer and usually appear in underlying modules that can be easily caught by unit testing~\cite{unittest}.
Third, MPE requires modifying DBMSs which might generate non-triggerable query plans. 
However, this case is rare. We found only one such case in practice.
It occurs only when one plan \textbf{always} costs more than another, such as scanning a table in parallel with only one worker is always worse than scanning a table non-parallel.
However, this case can also be viewed as a performance bug since the DBMS wastes efforts on generating meaningless query plans.

\begin{figure*}[th]
\centerline{\includegraphics[width=0.8\textwidth]{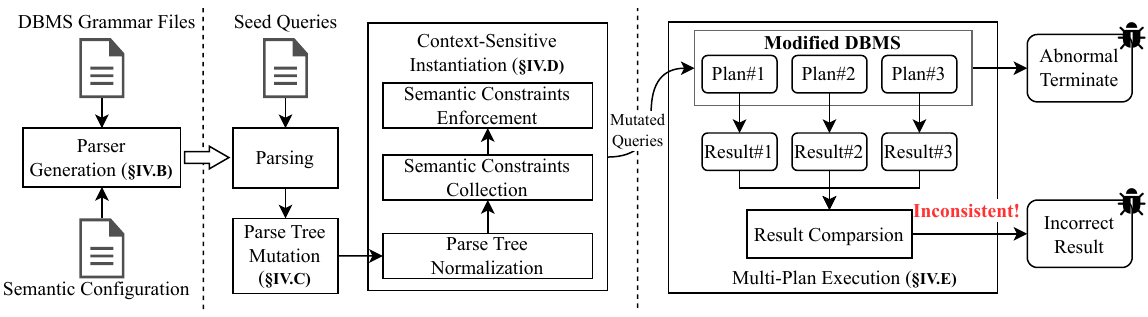}}
\caption{The overall architecture of our system.}
\label{fig:architecture}
% \vspace{-0.3cm}
\end{figure*}

%-------------------------------------------------------------------------------
\section{Design}
%-------------------------------------------------------------------------------

% In this section, we first illustrate the whole system design and then
% describe each component of the system. 

\subsection{Overall Design}

Our system aims at detecting both memory bugs and logic bugs in DBMSs.
Figure~\ref{fig:architecture} illustrates the overall architecture of our system.
The \textit{parser generation} (Section~\ref{ref_paerser_generation}) takes a semi-automatic way to generate SQL parsers before the testing. 
During the testing, it first performs a \textit{parse tree mutation} (Section~\ref{AST_mutator}) to mutate the test case while preserving syntactic correctness.
Because the mutation usually breaks the semantic validity of test cases, our system conducts a process of \textit{context-sensitive instantiation} (Section~\ref{subsec:context_sensitive_instantiation}) to improve the semantic correctness of the mutated queries.
At last, the mutated SQL queries will be executed by the DBMS.
The \textit{multiple-plan execution} (Section~\ref{sec:multi-testing}) hooks into the optimizer in DBMS
to execute all query plans (instead of the optimal one) and compare the returned results.
Any inconsistency between the results indicates a logic bug.
Similarly, a memory bug is detected if crashes or assertion failures occur.

\begin{figure*}[htp]
\centerline{\includegraphics[width=0.98\textwidth]{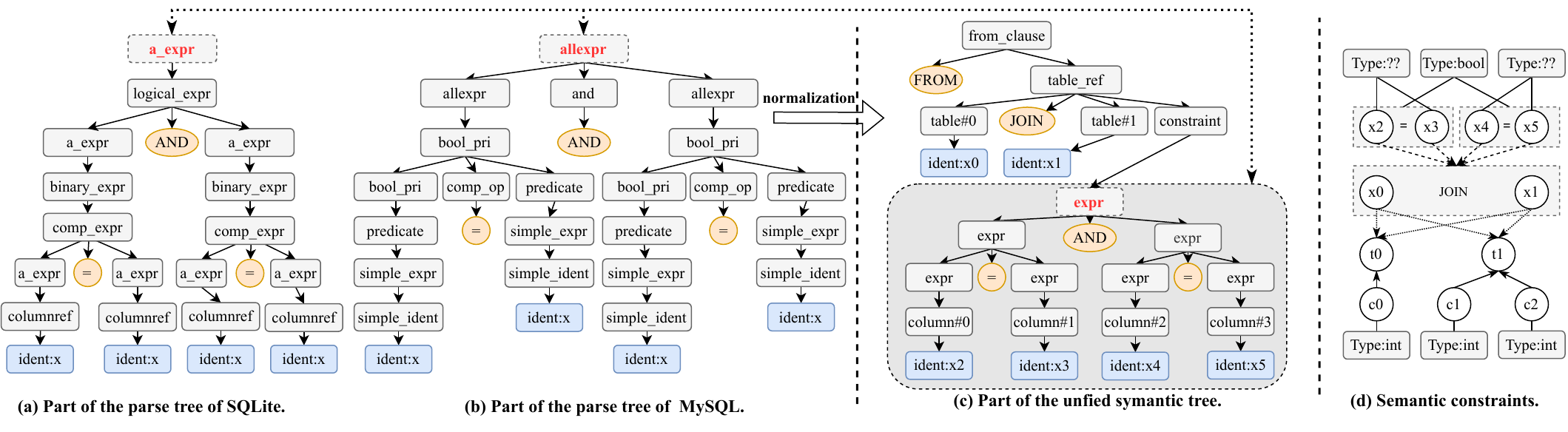}}
\caption{The process of context-sensitive instantiation in the example of Figure 2. Figure(a) shows part of the parse tree of SQLite for the SQL snippet \code{x=x AND x=x}.
Figure(b) shows the parse tree of MySQL for the same SQL snippet.
We normalize the parse tree of SQL snippet \code{x JOIN x ON x=x AND x=x} to a unified semantic tree in Figure(c).
In this way, we can perform analysis on the semantic tree regardless of differences in syntactic structure.
Figure(d) describes the semantic constraints we collected from the semantic tree.}
\label{fig:motivation_graph}
% \vspace{-0.3cm}
\end{figure*}

\smallskip
\noindent\textbf{Example}\tab
In the following, we use a SQL query (Figure~\ref{fig:multi-plan-example}) that will trigger a logic bug in SQLite to illustrate the flow of our system. 
Generating the test case shown in Figure~\ref{fig:multi-plan-example} takes several rounds of mutation on the initial seed.
We take the last round of mutation as an example.
The last round of mutation is performed on the statement \code{SELECT * FROM x JOIN x ON x=x AND x>i;}.
Suppose our system replaces the expression node \code{x>i} with the expression node \code{x=x}, 
which generates the statement \code{SELECT * FROM x JOIN x ON x=x AND x=x;}.

Our system then performs context-sensitive instantiation to replace the symbolic values with concrete values.
Specifically, it first normalizes the parse tree into a unified semantic tree shown in Figure~\ref{fig:motivation_graph}c to eliminate the differences in syntactic structure.
Then, it analyzes the semantic tree to collect semantic constraints of variables.
We utilize the randomized backtracking algorithm~\cite{rdfs} (also called randomized depth-first search) to find a random solution for all variables that satisfy all semantic constraints.
Specifically, we concrete variables from \code{x0} to \code{x5} in turn.
Variables \code{x0} and \code{x1} are table references that can refer to any table created by the previous \code{CREATE} statement.
Assuming we assigned \code{t0} to both \code{x0} and \code{x1}.
Then we concrete variable \code{x2}.
Because \code{x2}\textasciitilde\code{x5} are columns that refer to columns with unique names in joined table \code{x0 JOIN x1}, two candidate columns with the same name \code{c0} violates this constraint.
Thus, we backtrace to variables \code{x1} and concrete it with another random value \code{t1}.
In this case, instantiating \code{x2}\textasciitilde\code{x5} as any column satisfies all constraints.
If we randomly assign \code{c1}, \code{c0}, \code{c2}, and \code{c0} to \code{x2}, \code{x3}, \code{x4}, and \code{x5} respectively, we obtain a solution that satisfies all constraints.
% Picking the same table for \code{table\#0} and \code{table\#1} will result in unavailable values for \code{column\#0-3} (due to the ambiguous column names).
% Therefore, we pick \code{t0} and \code{t1} for \code{table\#0} and \code{table\#1}, respectively.
% Because \code{column\#0} and \code{column\#1} are the operands of a comparison expression, they should have the same data type.
% In particular, if we pick \code{c2} as the left operand, then the right operand
% cannot be \code{c1} since their types are different (we randomly pick \code{c0}).
At last, the concreted semantic tree will be translated to a SQL string as a new statement \code{SELECT * FROM t0 JOIN t1 ON c1=c0 AND c2=c0;}.
When the test case is executed, our system executes all four query plans one by one and
finds that three of them return an empty result, while the fourth plan returns a non-empty value.
This leads to the detection of the logic bug in the DBMS optimizer.

\subsection{Parser Generation}\label{ref_paerser_generation}
The syntactic rules of a SQL parser vary in different DBMSs.
A statement that can be executed by SQLite may be rejected by the parser of PostgreSQL.
To guarantee the mutated test case is syntactically correct, we need to have an accurate parser for each DBMS.

Implementing an accurate parser requires non-negligible effort.
To save manual effort, we propose a semi-automatic approach to generate the parser.
Our core insight is that almost all DBMSs leverage a standardized format called Backus Normal
Form (BNF)~\cite{knuth1964backus} to describe the grammar of the supported SQL queries.
By extracting and analyzing the BNF rules, we can construct an accurate parser.
Specifically, the parser generator extracts syntactic rules from the target DBMS grammar file.
Then, it uses syntactic rules and a semantic configuration to build the parser.
We explain the semantic configuration in parse tree normalization of Section~\ref{subsec:context_sensitive_instantiation}.
To minimize human intervention, the generated parser only implements the most fundamental functionality, i.e., generating a parse tree.
Some DBMS parsers do additional syntax checks beyond BNF rules, which is hard to be automatically ported.
We manually port these checks to the generated parsers.

\subsection{Parse Tree Mutation}
\label{AST_mutator}

% 看看Squirrel和TQS是怎么写的，这部分简单改写一下。
Our system mutates the SQL queries based on the parse tree. This can ensure the mutated queries are always syntactically correct.
Similar to \squirrel~\cite{zhong2020squirrel},
we also focus on structure mutation since it is more effective than data mutation.
Specifically, our system symbolizes all the variables in statements and concretes them after the mutation.

To mutate a test case, our system randomly picks up one node \textit{v} from
the parse tree and randomly fetches a new one \textit{w} with the same type from the mutation library. 
Then, it replaces \textit{v} (including its children) with \textit{w}.
The mutation library is a dictionary where the key is the node type and the value is a list of distinct parse trees rooted at nodes of that type.
Our system accepts DBMSs' official test cases to initialize the mutation library.
It is worthwhile noting that the mutation strategy is not a contribution of our work,
 as it is basically the same as the previous work~\cite{zhong2020squirrel}.

\subsection{Context-Sensitive Instantiation}
\label{subsec:context_sensitive_instantiation}

After the parse tree mutation, we need to analyze the SQL statement to collect semantic constraints and concretize all variables in mutated queries.
Building a general analysis module for SQL is challenging.
First, while DBMS internally has some capabilities to analyze a SQL query, the query analysis component is typically deeply embedded inside DBMS and is hard to extend.
Furthermore, each DBMS has its own analysis logic, specific to its own SQL dialect.

To address these challenges, context-sensitive instantiation first normalizes the parse tree to a unified semantic tree to make our method general to different DBMSs.
% Context-sensitive instantiation consists of three steps.
% First, it normalizes the parse tree to a unified semantic tree to make our method general to different DBMSs.
Then, it performs context-sensitive analysis on the semantic tree to get semantic information and infer semantic constraints.
Finally, it utilizes constraints to guide the instantiation of variables.

\smallskip
\noindent\textbf{Step I: Parse Tree Normalization}\tab
For a given SQL statement, the parse trees are typically completely different (Figures~\ref{fig:motivation_graph}a and \ref{fig:motivation_graph}b).
To make the semantic analysis general, we normalize the parse tree into a unified semantic tree.
Our key observation is that the parse tree has two types of nodes.
The first type refers to the SQL clauses and the second type is highly correlated to the detailed implementation of parsers in different DBMSs.
Since the first type of nodes preserves full semantic information, the unified semantic tree only catches all SQL clauses while ignoring the trivial implementation-dependent nodes.
As a standard interop format, the semantic tree can be 
1) inspected to collect semantic constraints 
2) modified to rewrite the query 
3) rendered back into a SQL dialect (as a string) without worrying about parsing or integration with different SQL dialects. 
During parsing, the parser labels the semantic types of these nodes according to the semantic configuration, which defines the mapping rule between SQL clauses and the parse node types. 
Each semantic type corresponds to a SQL clause.

\begin{figure}[t]
\centerline{\includegraphics[width=0.5\textwidth]{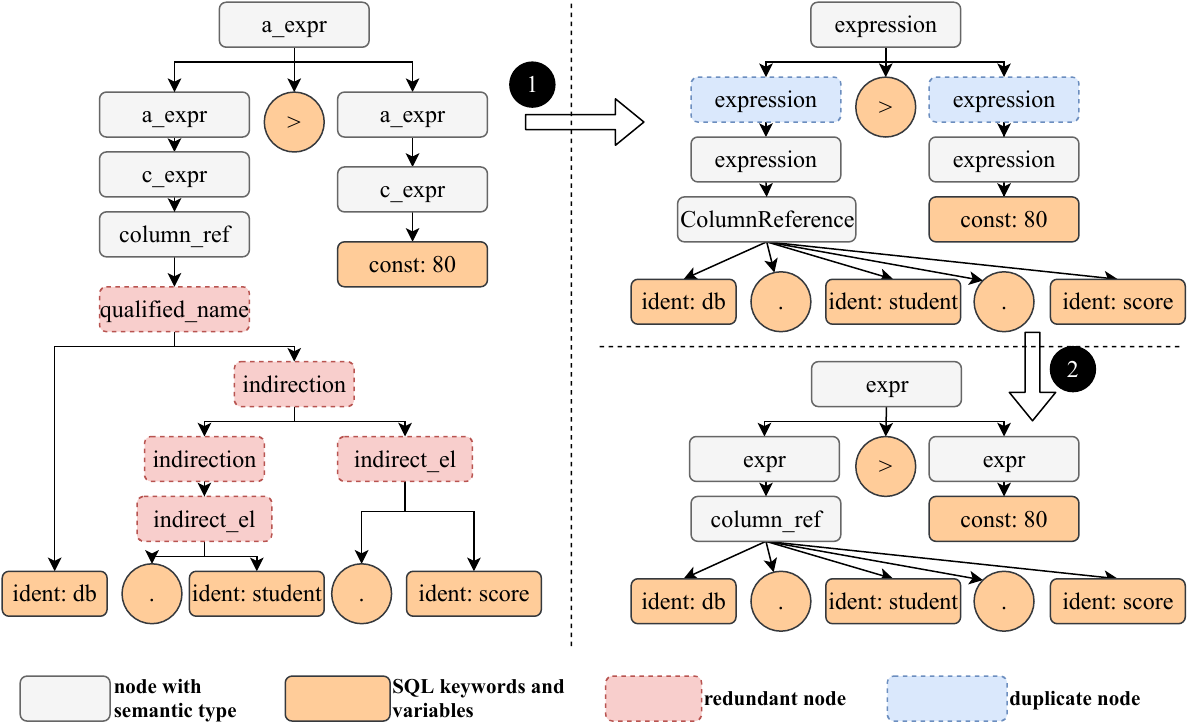}}
\caption{An example to show the process of translating a parse tree to a semantic tree. \ding{182} We remove redundant nodes, including \code{qualified\_name}, \code{indirection}, and \code{indirect\_el}.
We map the other nodes to the corresponding semantic nodes according to the user definition.
\ding{183} Then, We merge the duplicate semantic nodes in the semantic tree.}
\label{fig:tree_nomoralization}
% \vspace{-0.3cm}
\end{figure}

The parse tree normalization consists of two steps.
First, we convert parse tree nodes with semantic types into corresponding semantic nodes to build the semantic tree.
Then, we remove the duplicate semantic nodes in the semantic tree. 
We consider a node to be a duplicated one if it has one and only one child node of the same semantic type as its own.
Such duplication is mainly caused by semantic inclusion between different types of parse nodes.
For instance, in Figure~\ref{fig:tree_nomoralization}, 
the \code{c\_expr} represents the primary expression like a single column or a constant, while the \code{a\_expr} can also be a comparison expression.
The semantic type of \code{a\_expr} and \code{c\_expr} are both expressions, but the semantic of \code{c\_expr} is fully covered by \code{a\_expr}.

\smallskip
\noindent\textbf{Step II: Semantic Constraints Collection}\tab
Algorithm~\ref{alg:semantic_analysis} shows our semantic constraints collection algorithm.
For each SQL statement in the test case, our system traverses the semantic tree in post-order 
 and performs analysis on each node during the traversal.
The analysis gets the matched pattern of the SQL clause by analyzing the sequences of the child nodes.
The official documents list all possible patterns of SQL clauses and their semantic information.
Thus, we can get semantic information from the matched pattern and infer semantic constraints.
% For DBMS-specific clauses, we can easily extend and customize the analysis method on them.

\begin{algorithm}[t]
\scriptsize
\SetAlgoSkip{}
\caption{Algorithm of semantic analysis.}
\label{alg:semantic_analysis}
\LinesNumbered
\DontPrintSemicolon
\KwIn{SemanticTrees: the root node of the semantic tree of an SQL statement}
\KwOut{ConstraintSet: all semantic constraints of the current statement
        % contextInfo: context information of current statement
        % (e.g. the created table and its columns)
        }
\BlankLine
\SetKwProg{Fn}{Function}{:}{}
\SetKwData{constraints}{\scriptsize constraints}
% \SetKwData{contextInfo}{contextInfo}
% \SetKwData{context}{context}
\SetKwData{SemanticTree}{\scriptsize SemanticTree}
\SetKwData{SemanticNode}{\scriptsize SemanticNode}
\SetKwData{ConstraintSet}{\scriptsize ConstraintSet}
\SetKwData{MatchPattern}{\scriptsize MatchPattern}
\SetKwFunction{CollectConstraints}{\scriptsize CollectConstraints}
\SetKwFunction{PostOrderTraverse}{\scriptsize PostOrderTrav}
\SetKwFunction{Analyze}{\scriptsize Analyze}
\SetKwFunction{FindPattern}{\scriptsize FindPattern}
\SetKwFunction{Getconstraint}{\scriptsize Getconstraint}
% \SetKwFunction{ContextUpdate}{ContextUpdate}
% \SetKwFunction{GetContext}{GetContext}

\Fn{\CollectConstraints{\SemanticTree}} {
    \ConstraintSet$\gets$ vector()\;
    % \contextInfo$\gets$ vector()\;
    \PostOrderTraverse{\SemanticTree, \ConstraintSet}\;
    % \contextInfo$\gets$ \SemanticTree.context\;
    \KwRet{\ConstraintSet}
    % \KwRet{\constraints, \contextInfo}
}

\Fn{\PostOrderTraverse{\SemanticNode,\ConstraintSet}} {
    \tcp{children of the semantic node include semantic node, SQL keyword, and variables}
    \For{each child C in \SemanticNode.children}{
        \If{C is semantic node}{
            \PostOrderTraverse{C}
        }
    }
    \constraints $\gets$ \Analyze(C) %\tcp*[l]{invoke function Analyze after all children have been analyzed}
    \ConstraintSet.update(\constraints)\;
}

\Fn{\Analyze{\SemanticNode}} {
    \MatchPattern$\gets$ \FindPattern{\SemanticNode.children} \tcp*[l]{find the matching pattern of the SQL clause}
    \constraints$\gets$ \Getconstraint{\MatchPattern}\;
    % \context$\gets$ \GetContext{\MatchPattern}\;
    % \ContextUpdate{\context}\;
    \KwRet{\constraints}
}
\end{algorithm}

\begin{table*}[th]
\footnotesize
\caption{Semantic constraints. It explains the semantic constraints with simple examples.} % title of Table
\centering % used for centering table
\setlength{\tabcolsep}{0.48mm}{
\begin{tabular}{l|l|l|l} % centered columns (4 columns)
\toprule %inserts double horizontal lines
Constraint Type & Example SQL & Example Constraint & Explain \\ 
\hline
Variables type & {\scriptsize SELECT * FROM x1 ORDER BY i1;} & ${\scriptstyle Type(i1)=ColumnId}$ & i1 is a constant with type ColumnId \\
Data type & {\scriptsize SELECT x1+1 FROM x2;} & ${\scriptstyle DataType(x1)=numeric}$ & The data type of column x1 should be numeric \\
Name & {\scriptsize SELECT x1 FROM x2;} &  ${\scriptstyle UNIQUE(x1\in x2)}$ & x1 is a column with unique name in table x2 \\
Attribute & {\scriptsize INSERT INTO x1(x2) VALUES (i1);} & ${\scriptstyle Attr(x2)!=GENERATED}$ & x2 should not be a generated column \\
Value & \scriptsize SELECT * FROM x1 ORDER BY i1; & ${\scriptstyle i1\in\{1, 2, ..., Size(x1)\}}$ & i1 is an integer up to the number of columns of table x1 \\
Dependency & {\scriptsize SELECT x1 FROM (SELECT x2, x3 FROM x4) x5;} & ${\scriptstyle x1 \in \{x2, x3\}}$ & x1 could be column x2 or x3 \\
Distinct & \scriptsize INSERT INTO x1(x2, x3) ...; & ${\scriptstyle DISTINCT[x2, x3]}$ &  x2 and x3 should be two different column \\
Composite & \scriptsize SELECT (x1, x2) IN (TABLE x3) FROM x4; & ${\scriptstyle (DataType(x1), DataType(x2))) \in}$   & The data type of x2 and x3 should be \\
& & ${\scriptstyle \{DataType(x3,1), DataType(x3,2)\}}$ & the same as the first two columns in table x3. \\
\bottomrule
\end{tabular}}
\label{tab:semantic_constraint}
% \vspace{-0.3cm}
\end{table*}

Table~\ref{tab:semantic_constraint} lists all kinds of semantic constraints that we classify from the DBMSs' official grammar document.
The first five types of semantic constraints are constraints on the properties of variables.
The dependency and distinct constraints are constraints between different variables.
The composite constraint refers to the constraints on data consisting of multiple variables.

We use an example to better illustrate the process of analyzing a statement.
Figure~\ref{fig:comparsion-instantiation}b (on page~\pageref{fig:comparsion-instantiation}) shows
 the semantic analysis process on the \code{SELECT} statement.
We conduct analysis on semantic nodes (SQL clauses) from \normalsize{\textcircled{\scriptsize{1}}}\normalsize\enspace to \normalsize{\textcircled{\scriptsize{14}}}\normalsize.
The first node is a \code{ColumnReference} clause, thus the type of \code{x1} is \code{ColumnRef}.
In the following, we skip the introduction of analysis of nodes \normalsize{\textcircled{\scriptsize{2}}\textcircled{\scriptsize{5}}\textcircled{\scriptsize{6}}\textcircled{\scriptsize{7}}\textcircled{\scriptsize{8}}}\normalsize\enspace since they are similar.
% Similarly, we can infer that the \code{i1} is a \code{Constant}.
Next, we analyze the node \code{Expression} which consists of a node \code{ColumnReference}, a SQL keyword '\code{+}', and a node \code{Constant}.
Apparently, this is an addition expression, and the \code{ColumnReference} and \code{Constant} are operands.
Therefore, the data type of operands, \code{x1} and \code{i1}, should be numeric.
% After we finish analyzing \code{SelectTarget},
%  we analyze \code{TableReference} x2 and x3 and then \code{ColumnReference} x4, x5.
% We get their type during these analyses.
Then \code{Expression \$1=\$2} represents a comparison expression which provides a new constraint that the data type of \code{x4} and \code{x5} should be the same (e.g. both are numeric).
The \code{JoinConstraints} requires a logic value from the condition \code{\$1}.
% If \code{\$1} is a variable (e.g \code{ColumnRef}), we will add a constraint ${DataType(\$1)=boolean}$.
% Otherwise, we will check its data type.
% If the check fails, we will report a semantic error and terminate the mutation.
% That's because such semantic error is typically caused by a semantic-blind mutation (i.e., replacing a comparison expression with an arithmetic expression) and cannot be fixed by the variable instantiation.
The \code{TableReference \$1 JOIN \$2 \$3} joins two tables with a condition \code{\$3}.
We build a table set containing two tables \code{x2} and \code{x3} and add the constraint that all \code{ColumnReference} under the context of the join condition \code{\$3} can only be the column of the joined table.
At last, we analyze the root node of this tree, i.e., the \code{SelectStmt} clause.
All \code{ColumnReference} by default could be any column in the table built by \code{FromClause}.

\begin{lstlisting}[language=sql,float=t,
label={list:unfixable-example},
belowskip=4pt,
numbers=left,
numbersep=2pt,
xleftmargin=1em, 
commentstyle=\color{codegreen},
backgroundcolor=\color{mygray},
basicstyle=\scriptsize\ttfamily,
numberstyle=\scriptsize\ttfamily,
keywordstyle=\color{purple},
% keywordstyle=\bfseries,
breaklines=true,
caption={An example of the mutation that leads to semantic violations, which cannot be fixed by instantiating variables.}]
-- Origin test case ---------
CREATE TABLE t0(c0 INT, c1 INT);
INSERT INTO t0 VALUES (1, 2);
SELECT c0 FROM t0;
-- Mutated test case --------
CREATE TABLE t0(c0 INT, c1 INT);
INSERT INTO t0(c0) VALUES (1, 2);
-- Error: Element Number Inconsistency
SELECT c0 FROM t0 CROSS JOIN t0;
-- Error: Ambiguous Column Names
-- Patched test case --------
CREATE TABLE t0(c0 INT, c1 INT);
INSERT INTO t0(c0) VALUES (1);
SELECT r0.c0 FROM t0 AS r0 CROSS JOIN t0 AS r1;
\end{lstlisting}

\smallskip
\noindent\textbf{Step III: Semantic Constraints Enforcement}\tab
Conceptually, enforcing semantic constraints can be viewed as constraint satisfaction problems (CSPs)~\cite{CSPs}, i.e., finding a random solution for all variables that satisfies all the constraints.
% Traditional solvers usually only return a fixed solution for a given set of constraints.
% The problem of efficiently sampling diverse solutions to complex constraints is still challenging. 
% To the best of our knowledge, no mature constrained sampling tool is currently available.
Constraint satisfaction problems on finite domains are typically solved using a form of search.
Our system employs a randomized backtracking algorithm~\cite{rdfs} to solve these constraints.
Initially, all variables are unassigned.
At each step, a variable is chosen and a randomly possible value is assigned to it and the satisfaction of the partial assignment is checked. In case of satisfaction, a recursive call is performed, otherwise the algorithm backtracks.
We find a random solution if all variables are successfully assigned.
% If all values have been tried, the constraints are unsolvable.

% The high-level idea is to randomly enumerate all possible dependencies for each variable until finding one determined dependency that satisfies all constraints.
% Specifically, we first instantiate variables that have no internal dependencies.
% Then we concrete variables that have no undetermined dependencies.
% During variable instantiation, we randomly choose one possible dependency that satisfies all constraints.
% If all possible dependencies of a variable fail to satisfy the constraints, we will re-instantiate the variables it depends on.
% For example, in Figure~\ref{fig:comparsion-instantiation} (on page~\pageref{fig:comparsion-instantiation}), \code{x2}, \code{x3}, and \code{x6}
% only depend on previously declared tables \code{t0} and \code{t1}, so we first instantiate them in turn by randomly selecting a referenced table.
% Then we concretize \code{x1}, \code{x4}, \code{x5} because all variables they depend on
% have been instantiated.
% Suppose we concretize both \code{x2} and \code{x3} with table \code{t1}, the joined table will have two columns with conflicting names.
% When concretizing identifier \code{x4}, the column dependency requires that \code{x4} should be a column with a unique name in the joined table, but no dependencies satisfy this constraint.
% In this case, we will re-instantiate \code{x3}, \code{x4}, and all variables depend on them.

The semantic constraints of the mutated statement may be unsolvable
 or the semantics cannot be fixed by instantiating variables.
To enhance the robustness of the instantiation,
 we will try to patch the SQL statement in these cases.
Listing~\ref{list:unfixable-example} shows such an example.
The mutation on the \code{INSERT} statement leads to an inconsistent elements error that cannot be fixed by instantiating variables.
Specifically, the mutated \code{INSERT} statement provides a column list \code{(c0)} for the table.
However, this column list provides only one column but the insert values supply two columns.
This will cause a semantic error.
Our system ensures the sizes of the related lists are consistent by resizing these lists.
Another mutation performed on the \code{SELECT} statement causes unsolved constraints.
The mutated \code{SELECT} statement replaces the single table with a joined table.
Because only one table is created in the previous statement, the joined table must be a self-join table.
Since the dependency constraints require identifier \code{c0} to be a uniquely named column in the joined table, this constraint cannot be solved.
To address this issue, we attach an alias to each table and insert the table alias name before the column name.

\subsection{Multi-Plan Execution}\label{sec:multi-testing}
To enable logic bug detection, we employ MPE in our system.
MPE focuses on detecting logic bugs in the planner and executor because these components are the most complex ones and have not been well-tested. 
In DBMS, a SQL query will be translated to multiple query plans. Only the optimal query plan will be executed by default.
Our system hooks this process to \textit{execute all the query plans for a \code{SELECT} statement to detect the incorrect one(s) that may cause incorrect result}.
Currently, we choose SELECT statements to demonstrate MPE, aligned with the setting in previous works \cite{rigger2020detecting,rigger2020finding,rigger2020testing,liang2022detecting,tang2023detecting},
 since other statements typically lack returning results to be checked. 

Adopting MPE to DBMSs is not straightforward.
The first challenge is how to make DBMS execute all query plans with
a small modification.
% DBMS typically employs two types of optimizers at the same time: cost-based optimizers and ruled-based optimizers.
% The cost-based optimizers typically rely on
Our key observation is that DBMS typically employs a plan choose function to compare the estimated cost of different 
query plans (including their sub-plans) and choose the best one.
Therefore, we can execute any query plan by hooking the plan choose function.
In addition, we added a loop before the entry function for query processing
 so that it can be executed multiple times with different plans until all plans are executed.
Such modification has two advantages.
First, it does not break the functionality of DBMS,
 as it only interferes with the process of choosing a query plan.
Second, it requires little domain knowledge and implementation effort in practice.
% The rule-based optimizer uses a pre-defined set of rules to determine whether to apply optimizations.
% Some of these can be disabled at run-time through the interface provided by the DBMS, others are hard-coded enabled.
% For optimizers with switches, we can run the test cases twice, once with optimizations left on and a second time with optimizations turned off.
% For hard-coded optimizers, we simply ignore them, i.e. leave them enabled, since disabling them may generate incorrect query plans or even crash the DBMS.

The second challenge is reducing the false positives caused by non-deterministic queries.
A SQL query could be non-deterministic.
The inconsistency between the results of different query plans does not necessarily indicate a bug.
We test a large subset of each DBMS's features and identify many features that may cause false positives.
We investigate the root cause of the factors leading to non-deterministic behaviors and catalog them into four types.

The first source of non-determinism comes from undefined data access order.
%, which is also the fundamental basis for optimization. 
Since the needed data can be fetched by accessing it in different ways and in different orders, the orders of intermediate and final results are non-deterministic.
In most cases, we sort the results to eliminate such non-determinism.
However, the non-deterministic cannot be resolved if the statement contains some specific features.
For example, the \code{LIMIT} clause results in only part of the records being fetched. 
Therefore, different data access orders may lead to inconsistencies in the records that are finally fetched.
The second category is caused by non-deterministic functions or variables such as \code{random()} and \code{CURRENT\_TIME}. 
Another common cause is that decimal calculations may lose precision, which is common in DBMSs' built-in statistical analysis functions.
At last, some undefined behaviors depend on the dynamic execution context.
Thus, different query plans may lead to different contexts and ultimately result in inconsistent results.

To solve this issue, we mark the statements containing non-deterministic features during context-sensitive analysis.
For these marked statements, our system skips the result comparison to avoid false positives.
That is to say, we can still detect memory bugs from non-deterministic queries.

\subsection{Prototype Implementation}
\label{subsec:implementation}

\begin{table}[t]
\caption{The lines of code of different components}
\centering
\setlength{\tabcolsep}{0.48mm}{
\begin{tabular}{l c c}
\toprule
Component & Language & Lines of code  \\
\hline
Parser generator & Python & 1917  \\
Parse-tree mutator & C++ & 849 \\
Context-sensitive instantiation & C++ & 12,476 \\
Result comparison & C++ & 909 \\
Fuzzer & C & 5,241 \\
Other & - & 798 \\
\hline
Total & - & 22,190 \\
\bottomrule
\end{tabular}}
% \vspace{-0.3cm}
\label{table_codesize}
\end{table}

We have implemented a prototype system called \sysname and applied it to three widely-used DBMSs, including SQLite, PostgreSQL, and MySQL.
% To the best of our knowledge, \sysname is the first coverage-guided~\cite{zhou2020zeror} DBMS fuzzer for optimization bugs detection.
Our system consists of 21.9k lines of code (LoC) in total.
Table \ref{table_codesize} summarizes each of the components.
Our system is built on top of AFL 2.56b~\cite{AFL}. 
% Specifically, we replace its mutator with ours and integrate QueryPlan Sanitizer
% into it.

SQL language in DBMS contains more than 100 different types of clauses.
Some of these clauses are standard SQL clauses, and some are dialects.
Each clause usually has multiple patterns corresponding to different semantics.
For example, the \code{ColumnConstraint} clause defines the attribute of a column, which contains nine different patterns, e.g., \code{PRIMARY KEY}, \code{GENERATED}, and \code{FOREIGN KEY}. 
Our prototype implements semantic analysis for the primary patterns of the 45 most frequently used clauses.
42 of these are shared by the three target DBMSs and the others are DBMS-specific.
We understand that the unsupported patterns and clauses may lead to incomplete semantic constraints, therefore lowering the semantic validity probability of statements that have these clauses.
But our system still captures more semantic constraints than existing ones.

\section{Evaluation}

In this section, we answer the following research questions to show the advantages and effectiveness of \sysname.
\begin{itemize}
    % \item \textbf{Benefits of parser generation.} How effective is the parser generation approach in reducing manual efforts and improving the parser accuracy? (\ref{bpg})
    \item \textbf{Effectiveness of detecting bugs in real-world DBMSs.} How effective is \sysname's approach in discovering new bugs in real-world production-level DBMSs (Section~\ref{sec:real-world-bugs})?
    \item \textbf{Generating valid queries.} How effective is \sysname in generating valid queries
    (Section~\ref{sec:query_validity_test})?
    \item \textbf{Comparison with existing tools.} Can \sysname outperform previous DBMS testing tools
    (Section~\ref{sec:comparions-with-other-tool})?
    \item \textbf{Benefits of the proposed two key techniques.} How effective are context-sensitive instantiation and MPE in exploring new paths and detecting bugs (Section~\ref{sec:unit_test})?
\end{itemize}

\smallskip
\noindent
\textbf{Experimental Setup}\tab
We perform all the experiments on a computer with Ubuntu 18.04 system, Intel Core i7-7700, and 32GB RAM. We enlarge the bitmap to 512K bytes to mitigate path collisions~\cite{gan2018collafl}. 
For new bug detection, we started detecting bugs in their latest release version, which was SQLite 3.32.0, PostgreSQL 14.2 and MySQL 8.0.25, and switched to the newly released version when available.
For comparison evaluation, we aligned with the setting in previous works which choose the older version of DBMSs, i.e. SQLite version 3.34.0, PostgreSQL version 14.2, and MySQL version 8.0.11.
% Since \sqlancer requires the particular SQLite version 3.34.0, we use this version for comparison.
% For other DBMSs, we use the latest version, i.e., PostgreSQL version 14.2 and MySQL version 8.0.29.
% For all three DBMSs, we started detecting bugs in their latest release version, which was SQLite 3.32.0, PostgreSQL 14.2 and MySQL 8.0.25, and switched to the latest trunk version.
%We use the latest release version of DBMSs in bug detection (such as SQLite 3.35.4, PostgreSQL 14.3, and MySQL 8.0.29).

% part1: 
\subsection{Detecting Bugs in Real-world DBMSs}\label{sec:real-world-bugs}
As shown in Table~\ref{tab:bug_list_table}, 
across intermittent runs during a 20-month period of development, \sysname successfully discovered 50 unique bugs, including 41 memory bugs and 9 logic bugs.
22 of these bugs were triggered by non-optimal query plans.
We consider that a bug is unique based on the official assigned ID. 
Multiple bugs may fall under the same CVE ID. 
At the time of writing, all bugs have been confirmed, and 35 of them have been fixed with 11 CVEs assigned.
\textit{The SQLite developers responded that many of the bugs we reported have been in the code for many years and no fuzzers have ever run across it, despite SQLite being heavily tested~\cite{sqlitest} and used in literally millions of applications.}

\begin{table}[t]\footnotesize
\caption{Detected bugs list. \sysname detected 50 unique bugs, including 19
crashes, 9 logic bugs and 22 assertion failures. Only MySQL developers assign severity for bugs submitted via the bug reporting page.}
\centering
\setlength{\tabcolsep}{1mm}{
\begin{tabular}{l l l l l l} 
\multicolumn{6}{c}{\textbf{LB}: Logic Bugs \textbf{CS}: Crashes \textbf{AF}: Assertion Failure}  \\
\multicolumn{6}{c}{\textbf{S1}: Critical \textbf{S2}: Serious \textbf{S3}:Non-critical \textbf{S6}:Debug builds} \\
\toprule
\textbf{ID} & \textbf{Type} & \textbf{Function} & \textbf{Status} & \textbf{Severity} & \textbf{Reference} \\
\hline
\rowcolor{gray!40}
\multicolumn{6}{l}{\textbf{SQLite}}  \\
1   & CS     &   fts3 and snippet          & Fixed  & - & \scriptsize cve-2020-23568    \\
2   & CS     &   fts3 and matchinfo        & Fixed  & - & \scriptsize cve-2020-23569    \\
3   & CS     &   fts3 and ALTER            & Fixed  & - & \scriptsize cve-2020-23570    \\
4   & CS     &   multi-or covering index   & Fixed  & - & \scriptsize 376e07            \\
5   & CS     &   LIKE and OR optimizer     & Fixed  & - & \scriptsize 84fe52            \\
6   & CS     &   UNION ALL                 & Fixed  & - & \scriptsize 2aa354            \\
7   & LB     &   equivalence transfer      & Fixed  & - & \scriptsize 13976a            \\
8   & LB     &   IS NOT NULL AND expr      & Fixed  & - & \scriptsize 6a1424            \\
9   & LB     &   GROUP BY NULL             & Fixed  & - & \scriptsize 0094d8            \\
10  & LB     &   type affinity             & Fixed  & - & \scriptsize 0c437a            \\
11  & LB     &   expression tree           & Fixed  & - & \scriptsize 6a1424            \\
12  & AF     &   query flatten             & Fixed  & - & \scriptsize a97bbd            \\
13  & AF     &   aggregate queries         & Fixed  & - & \scriptsize d49628            \\
14  & AF     &   NEVER()                   & Fixed  & - & \scriptsize bfb7ce            \\
15  & AF     &   window function           & Fixed  & - & \scriptsize 9d5aa9            \\
\hline
\rowcolor{gray!40}
\multicolumn{6}{l}{\textbf{PostgreSQL}} \\
1   & CS     &  empty column value         & Fixed  & - & \scriptsize \#17477           \\
2   & AF     &  table alias                & Fixed  & - & \scriptsize \#17480           \\
\hline
\rowcolor{gray!40}
\multicolumn{6}{l}{\textbf{MySQL}} \\
1    & CS  &  parser                                     & Fixed     & S1 & \scriptsize cve-2021-2427       \\
2    & CS  &  parser                                     & Fixed     & S1 & \scriptsize cve-2022-21303      \\
3    & CS  &  storage                                    & Fixed     & S1 & \scriptsize cve-2022-21304      \\
4    & CS  &  val\_int                                   & Fixed     & S1 & \scriptsize cve-2022-21640      \\
5    & CS  &  LEX                                        & Fixed     & S1 & \scriptsize cve-2022-39400      \\
6    & CS  &  find\_item                                 & Fixed     & S1 & \scriptsize cve-2022-21638      \\
7    & CS  &  fix\_semijoin\_strategies                  & Fixed     & S1 & \scriptsize cve-2022-21638      \\
8    & CS  &  make\_active\_options                      & Fixed     & S1 & \scriptsize cve-2022-39400      \\
9    & CS  &  table\_contextualize                       & Fixed     & S1 & \scriptsize cve-2022-21528      \\
10   & CS  &  query\_block\_is\_recursive                & Fixed     & S2 & \scriptsize cve-2023-21917      \\
11   & CS  &  create\_tmp\_table                         & Fixed & S1 & \scriptsize 107825              \\
12   & CS  &  WITH RECURSIVE                             & Confirmed & -  & \scriptsize S1649226            \\
13   & LB  &  materialization\_lookup                    & Confirmed & S3 & \scriptsize 107576              \\
14   & LB  &  semi\_and\_left\_join                      & Confirmed & S3 & \scriptsize 107585              \\
15   & LB  &  indexed\_materialization                   & Confirmed & S3 & \scriptsize 107629              \\
16   & LB  &  UNION                                      & Confirmed & -  & \scriptsize S1651202            \\
17   & AF  &  CREATE VIEW UNION                          & Fixed     & S6 & \scriptsize 107471              \\
18   & AF  &  cond\_bool\_func                           & Fixed     & S6 & \scriptsize 107578              \\
29   & AF  &  val\_real                                  & Fixed     & S6 & \scriptsize 107638              \\
20   & AF  &  join\_read\_const\_table                   & Fixed     & S6 & \scriptsize 107681              \\
21   & AF  &  ft\_init\_boolean\_search                  & Fixed     & S6 & \scriptsize 107733              \\
22   & AF  &  tmp\_table\_field\_type                    & Fixed     & S6 & \scriptsize 107826              \\
23   & AF  &  optimize\_aggregated\_query                & Fixed     & S6 & \scriptsize 107647              \\
24   & AF  &  ha\_index\_init                            & Confirmed & S6 & \scriptsize 107636              \\
25   & AF  &  val\_decimal                               & Confirmed & S6 & \scriptsize 107660              \\
26   & AF  &  select\_in\_like\_transformer              & Confirmed & S6 & \scriptsize 107661              \\
27   & AF  &  create\_ref\_for\_key                      & Confirmed & S6 & \scriptsize 107663              \\
28   & AF  &  recalculate\_lateral\_deps                 & Confirmed & S6 & \scriptsize 107704              \\
29   & AF  &  fix\_outer\_field                          & Confirmed & S6 & \scriptsize 107719              \\
30   & AF  &  join\_read\_key\_unlock\_row               & Confirmed & S6 & \scriptsize 107722              \\
31   & AF  &  having\_as\_tmp\_table\_cond               & Confirmed & S6 & \scriptsize 107723              \\
32   & AF  &  add\_key\_field                            & Confirmed & S6 & \scriptsize 107768              \\
33   & AF  &  mdl\_request\_init                         & Confirmed & S6 & \scriptsize 108237              \\
\bottomrule
\end{tabular}
}
\label{tab:bug_list_table}
% \vspace{-0.3cm}
\end{table}

\begin{table}[t]
\caption{The percentage of semantic correctness of query validation. } % title of Table
\centering % used for centering table
\setlength{\tabcolsep}{0.8mm}{
\begin{tabular}{c c c c} % centered columns (4 columns)
% \hline %inserts double horizontal lines  % inserts table
\toprule
& SQLite & PostgreSQL & MySQL \\
%heading
\hline % inserts a single horizontal line
Squirrel & 24,792(53.1\%) & 5,869(17.9\%) & 16,283(15.1\%) \\
SQLRight & 28,710(61.5\%) & 8,508(26.0\%) &  35,012(32.5\%) \\
Kangaroo & 32,880(70.4\%) & 16,269(49.7\%) & 46,763(43.4\%) \\
\hline
Total & 46,672 & 32,753 & 107,693 \\
% Our tool & 45 & 300 & 556 \\ [1ex] % [1ex] adds vertical space
% \hline %inserts single line
\bottomrule
\end{tabular}}
\label{query_validity_table}
% \vspace{-0.3cm}
\end{table}

% \begin{table}[t]
% \caption{The summary of detected bugs. } % title of Table
% \centering % used for centering table
% \setlength{\tabcolsep}{0.8mm}{
% \begin{tabular}{c c c c c c} % centered columns (4 columns)
% % \hline %inserts double horizontal lines  % inserts table
% \toprule
% \textbf{Bug Type} & \textbf{Status} & SQLite & PostgreSQL & MySQL \\
% %heading
% \hline % inserts a single horizontal line
% \multirow{2}{*}{Memory} & Confirmed & 10 & 2 & 29 \\
%                         & fixed     & 10 & 2 & 18 \\
% \hline
% \multirow{2}{*}{Logic}  & Confirmed & 5  & 0 & 4  \\
%                         & fixed     & 5  & 0 & 0  \\
% \hline
% \textbf{Total} & \textbf{50} & \textbf{15} & \textbf{2} & \textbf{33} \\
% \bottomrule
% \end{tabular}}
% \vspace{-0.3cm}
% \label{tab:bug_summary}
% \end{table}

% part2: semantic guide analyze
\subsection{Generating Valid Queries}\label{sec:query_validity_test}
% A query that is semantically correct is a prerequisite for detecting optimization bugs. 
% Compared with Squirrel, Kangaroo further improves semantic analysis and builds richer data dependencies to improve the effectiveness of queries.
To measure how effectively our system generates valid SQL queries, we compare query validation of \sysname, \squirrel, and \sqlright.
Due to the randomness of mutations and the large gap between instantiating different SQL queries,
 comparing the validity rate of the generated test case cannot well illustrate the effectiveness of context-sensitive instantiation. 
To provide a fair comparison, we directly feed the same inputs to the query instantiation modules of these systems and verify the validity of their outputs.
To obtain a suitable benchmark, we gather SQL queries from official DBMS unit tests and strip the concrete value of all variables in these queries.
% Since Squirrel can only parse a small subset of SQL statements, we port our parser to Squirrel to eliminate the disadvantage caused by the failure of parsing. 

Table~\ref{query_validity_table} shows our evaluation results. 
\sysname achieves the highest semantic correctness in all three DBMSs.
This demonstrates that context-sensitive instantiation outperforms previous work.
\sysname still cannot achieve full semantic validity for two reasons.
First, context-sensitive instantiation does not consider dynamic constraints.
Second, the static constraints are incomplete since our prototype only implements analysis for primary patterns of major SQL clauses (Section~\ref{subsec:implementation}).

\sysname achieves a smaller improvement on SQLite than other DBMSs.
This is caused by the fact that SQLite's attempt to follow Postel's law~\cite{postel1981transmission}, which means it would accept unusual inputs and do its best to execute them rather than raising an error.
For example, \code{SELECT 1+'a'} is an acceptable SQL query on SQLite but will cause an operand type mismatch error in PostgreSQL.
% For the same result, all fuzzing systems achieve the best result in SQLite.
% Therefore, SQLite has minimal room for improvement.

\begin{figure*}[t]
\centerline{\includegraphics[width=0.98\textwidth]{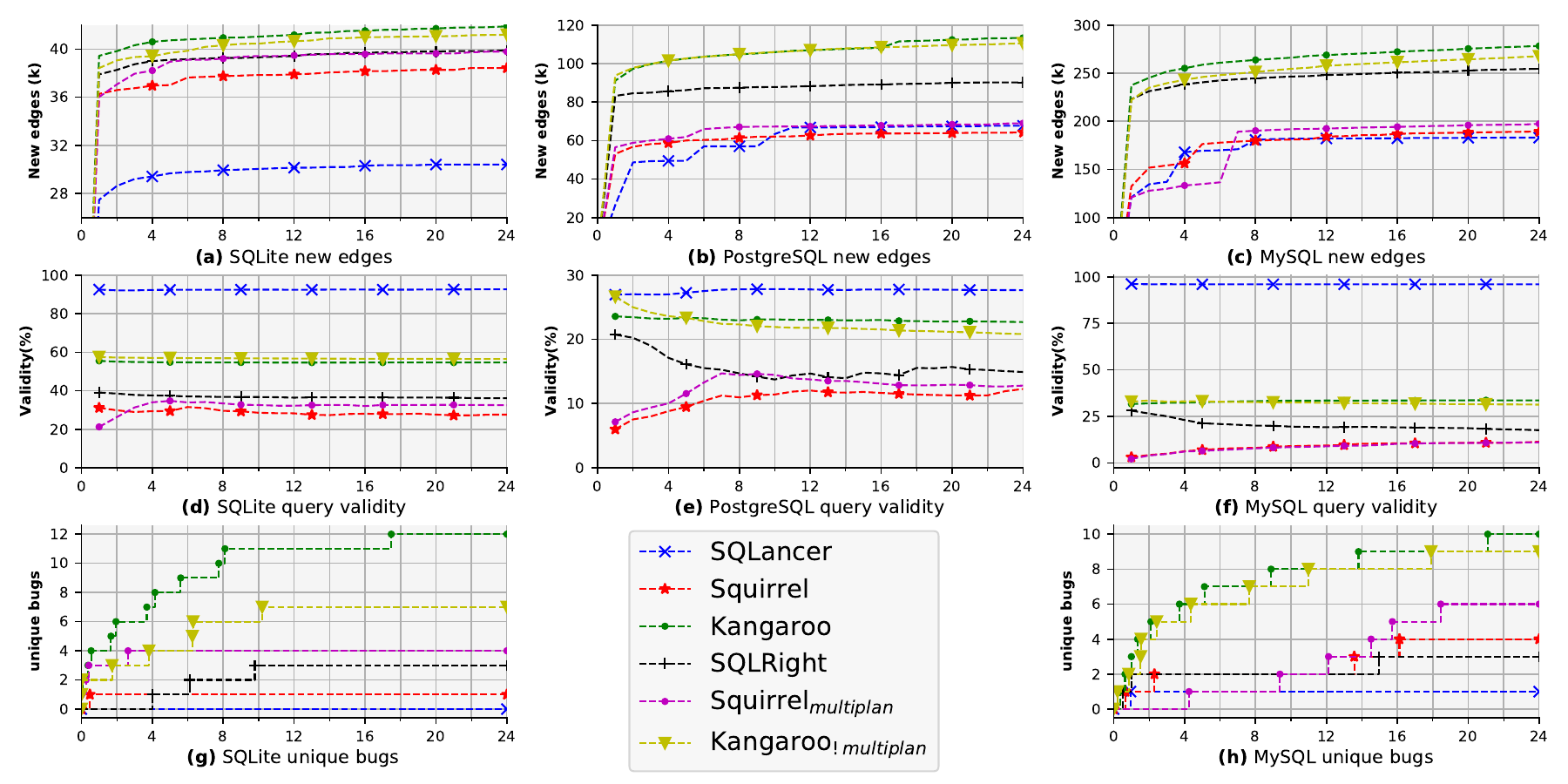}}
\caption{Comparison with existing tools. It also illustrates the contributions of context-sensitive instantiation and multi-plan execution. We exclude the results of detected bugs in PostgreSQL since only \sysname found one memory bug. We use NoREC for \sqlancer and \sqlright. Since \sqlancer does not implement NoREC for MySQL, we use TLP instead. }
% \vspace{-0.3cm}
\label{fig:comparison}
\end{figure*}

\subsection{Comparisons with Existing Tools}\label{sec:comparions-with-other-tool}

We compare \sysname with three state-of-the-art and open source systems: \squirrel, \sqlancer, and \sqlright.
% The comparison is fully aligned with the setting in previous works.
Similar to the previous ones, we also select SQLite, PostgreSQL, and MySQL for evaluation. 
We feed the same test cases to \squirrel, \sqlright, and \sysname as the initial corpus and provide the same queries used to initialize the mutation library.
\sqlancer is a generate-based tool that does not require any initial inputs.
We launch five fuzzing instances for each system and run each instance for 24 hours.
We report the average result except for the bug number. 
We collect all bug reports from the five fuzzing instances as the final result and then count their first occurrence time for each unique bug.
Figure~\ref{fig:comparison} shows the evaluation result.
% including the number of new edges, the query validity, and the total number of unique bugs.

\noindent
\textbf{Detected Unique Bugs}\tab
As indicated in Figure~\ref{fig:comparison}gh, \sysname outperforms existing tools on all three DBMSs.
For SQLite, \sysname found 12 unique bugs, including eight memory bugs and four logic bugs.
\sqlright found three bugs, one of which is a logic bug.
\squirrel found only one bug which is also covered by \sysname.
For PostgreSQL, only \sysname found one bug.
This is not surprising as previous works have revealed that PostgreSQL is
more robust than most other DBMSs.
As for MySQL, \sysname found a total of ten bugs, including eight memory bugs and two logic bugs.
\sqlright found two memory bugs and one logic bug.
\squirrel and \sqlancer detected five and one memory bugs, respectively.
\sqlancer, the only generation-based tool, found the least bugs across all comparisons, demonstrating the advantage of the mutation-based method to detect DBMS bugs.

\noindent
\textbf{Explored New Edges}\tab
Figure~\ref{fig:comparison}abc shows that \sysname performs better than others on all three DBMS systems.
It explores 52\%, 44\%, and 14\% more edges than \sqlancer, \squirrel, and \sqlright on average, respectively.
Considering that MPE modifies only a few lines of code out of millions of lines of DBMS code, we believe its impact on the new edges explored is negligible.
This means the improvement is mainly due to context-sensitive instantiation.

\noindent
\textbf{Generated Valid Queries}\tab
We treat a query as invalid if DBMS reports any form of error during the execution.
As shown in Figure~\ref{fig:comparison}def, \sqlancer achieves the highest query validity.
This result is reasonable because \sqlancer follows very limited grammar rules to generate SQL statements.
% In other words, the statements it generated contain only a small subset of SQL features.
For example, \sqlancer does not support generating subqueries that are prone to semantic errors.
That makes it easier to generate valid statements but limits the diversity.
This could be the main reason that it explores the fewest paths among all fuzzers.
Benefiting from the richer semantic constraints due to context-sensitive instantiation, \sysname achieves a noticeably higher query validity than other mutation-based fuzzers.

% \begin{figure*}[th]
% \centerline{\includegraphics[width=0.98\textwidth]{Fig/oracle_compare.png}}
% \caption{Oracle Comparsion. We exclude the results of detected logic bugs in PostgreSQL since none of the systems detected logic bugs.}
% \label{fig:oracle_comparsion}
% \vspace{-0.3cm}
% \end{figure*}

\noindent
\textbf{Oracle Comparsion}\tab
% \subsection{Oracle Comparsion}\label{sec:oracle_compare}
% We compare MPE with all previous oracles to demonstrate its effectiveness in finding logic bugs.
To eliminate the benefit from context-sensitive instantiation,
We build \sysnamenospace$_{MPE}$ by replacing the context-sensitive instantiation with type-sensitive instantiation in \sysname,
 and compare \sysnamenospace$_{MPE}$ to \sqlrightnospace$_{NoREC}$ and \sqlrightnospace$_{TLP}$.
Doing an automatic comparison of the MPE and PQS is difficult because PQS requires a generation method to generate statements.
Considering that the generation of test cases is also one of the evaluation metrics of the oracle, we include \sqlancernospace$_{PQS}$ as the comparison target.
After 5 rounds of 24-hour testing, \sysnamenospace$_{MPE}$ finds three logic bugs in SQLite and two logic bugs in MySQL, and \sqlrightnospace$_{NoREC}$ finds one logic bug in SQLite and one in MySQL. Besides, all logic bugs found by NoREC are covered by MPE.
\sqlrightnospace$_{TLP}$ and \sqlancernospace$_{PQS}$ fail to detect any logic bug.

\subsection{Benefits of the Two Key Techniques}\label{sec:unit_test}

% \begin{figure*}[t]
% \centerline{\includegraphics[width=0.98\textwidth]{Fig/unit_test.png}}
% \caption{Contributions of context-sensitive instantiation and multi-plan execution.}
% \label{fig:contribute_of_factor}
% \end{figure*}

We conduct unit tests to understand the contribution of the two techniques in \sysname.
To understand the contribution of context-sensitive instantiation, we build \sysnamenospace$_{!multiplan}$ by disabling the MPE in \sysname.
% \sysname$_{!multiplan}$ can also be viewed as \squirrel embedded with context-sensitive instantiation.
Without the MPE, \sysname$_{!multiplan}$ focuses on memory bug detection.
We use \squirrel as a baseline since it performs better than \sqlancer in detecting memory bugs.
We also ported the MPE module into \squirrel, denoted \squirrelnospace$_{multiplan}$.
By comparing \squirrel with \squirrelnospace$_{multiplan}$, we can verify that MPE itself is effective.
We adopt the same strategy to measure the number of explored edges, the query validity rate, and the number of unique bugs.
Figure~\ref{fig:comparison} shows the evaluation results.

% Overall, Kangaroo$_{!validity}$ only disables the extra semantic constraints enforcement which means that it only provides a more precise parser than Squirrel$_{multiplan}$; Kangaroo further improves semantic correctness compared to Kangaroo$_{!validity}$ and get additional coverage feedback compared to Kangaroo$_{!feedback}$.

\smallskip
\noindent\textBF{Context-Sensitive Instantiation}\tab
% To understand the benefit of context-sensitive instantiation,
% we compare \sysnamenospace$_{!multiplan}$ to \squirrel and \sysname to \squirrelnospace$_{multiplan}$.
Context-sensitive instantiation directly improves the correctness level (Table~\ref{tab:testcase_levels}) of generated test cases, thereby greatly improving the efficiency of fuzzing.
As shown in Figure~\ref{fig:comparison}def, tools using \textit{context-sensitive instantiation} achieve significantly higher semantic correctness than other tools.
By comparing \squirrel and \squirrelnospace$_{multiplan}$, we found that existing type-sensitive mutations cannot take full advantage of MPE due to low semantic correctness.
With the help of context-sensitive instantiation, \sysname significantly explores more edges than \squirrelnospace$_{multiplan}$.
\sysname detects 8 more bugs than \squirrelnospace$_{multiplan}$ in SQLite and 4 more bugs in MySQL.

% 1. 在Squirrel的基础上引入context-sensitive instantiation, 能极大的提升fuzzing效率。 2.强调在Squirrel基础上引入MPE，虽然有效果。但是再引入context-sensitive instantiation能进一步大大提升效率。
% For instance, \sysnamenospace$_{!multiplan}$ achieve 1.3x-2.2x semantic correctness than \squirrel, and therefore explore 2x more new edges and found more bugs in SQLite and MySQL.
% As shown in Figure~\ref{fig:comparison}def, tools using \textit{context-sensitive instantiation} achieve significantly higher semantic correctness than other tools.

\smallskip
\noindent\textBF{Multi-Plan Execution}\tab
Comparing \squirrelnospace$_{multiplan}$ to \squirrel and \sysname to \sysnamenospace $_{!multiplan}$ reveals the effect of MPE applied to two different DBMS testing systems.
First, MPE enables logic bug detection.
It has an average of 16.3, 15.7, and 16.1 query plans in the SELECT statements of test cases for SQLite, PostgreSQL, and MySQL, respectively.
\squirrelnospace$_{multiplan}$ identified one logic bug in SQLite with the help of the MPE technique, and Kangaroo discovered three logic bugs in SQLite and one in MySQL.
Interestingly, MPE also improves the capability to find memory bugs.
\squirrelnospace$_{multiplan}$ captures four more memory bugs than \squirrel in SQLite and MySQL, and \sysname finds two more memory bugs than \sysnamenospace$_{multiplan}$.
Among 50 unique bugs detected by our system, 22 of them are first detected in non-optimal query plans which indicates the effectiveness of MPE.
Our observation is that some query plans are rarely executed by DBMS and therefore are less tested.
MPE can execute these less-tested query plans to find more bugs.
In addition, MPE also has a slight improvement in coverage (1.7\%$\sim$13.1\%), which we believe is also attributed to the exploration of rarely executed query plans.

% MPE单独使用效果相对较差, 这是因为无效语句无法采用MPE策略。Context-sensitive instantiation能更好的帮助MPE发挥其作用。

\section{Discussion}

% \noindent
% \textbf{False Positives}\tab
% The false positives of our system come from two aspects.
% The first is the non-deterministic behavior of DBMS. As described in Section~\ref{sec:multi-testing},
% our system maintains a list of SQL features that can cause queries to be non-deterministic.
% However, this list could be incomplete.
% % During semantic analysis, it records all the SQL features of the queries and decides whether to compare the results of multiple query plans based on the list.
% But in practice, during more than one week of testing, our system does not discover any new features that lead to false positives.
% The second is the NPE for DBMS. Our modification to the DBMS only changes the selection logic of the query plans without involving the generation and execution of the query plan, so it will not break the functionality of the DBMS theoretically.
% But the bug in our code can only result in abnormal termination or select an unexpected query plan (e.g. for a given query, we select a duplicate plan or miss some plans).
% However, none of these errors can incur a false positive.

\noindent
\textbf{PoC Generation}\tab
A test case triggering a bug on the modified DBMS will not work on the original DBMS if the bug can only be triggered by non-optimal query plans.
As a result, such test cases cannot directly serve as proof-of-concept (PoC)~\cite{PoC}.
Nonetheless, the bug-triggering query plan on the modified DBMS and the corresponding program execution path is sufficient to find the root cause of the bug.
To save the DBMS developers’ time and effort, we submit all bugs and the corresponding minimal PoC.
To build minimal PoC, we first try to remove statements or clauses in the test case and check whether the bug is still triggered.
Then, we manually adjust the minimal test cases to make it can be triggered on the unmodified DBMS.
We spend about 20 hours manually constructing PoC for all 22 bugs detected in non-optimal query plans.
We provide some insight about manually adjusting PoC in Appendix~\ref{sec:poc_generation}.
We leave the automatic generation of PoC as part of future works.
% We leave the automatic generation of PoC as future works.
% The PoC generation is out of our research scope, but it has a huge impact on bug reports to developers. 
% How to automatically generate the PoC is one of the future works.

\begin{table}[t]
\caption{Effort of adoption to DBMSs.} % title of Table
\centering % used for centering table
\setlength{\tabcolsep}{0.8mm}{
\begin{tabular}{c c c c} % centered columns (4 columns)
% \hline %inserts double horizontal lines  % inserts table
\toprule
Component(LOC) & SQLite & PostgreSQL & MySQL \\
%heading
\hline % inserts a single horizontal line
Semantic configuration & 102 & 109 & 132 \\
Parser & 52 & 161 & 63 \\
Context-sensitive instantiation & 106 & 135 & 259 \\
DBMS modification & 248 & 271 & 191 \\
\hline
Total Person-hours & 32 & 40 & 45 \\
% Our tool & 45 & 300 & 556 \\ [1ex] % [1ex] adds vertical space
% \hline %inserts single line
\bottomrule
\end{tabular}}
\label{tab:addoption_effort}
% \vspace{-0.3cm}
\end{table}

\smallskip
\noindent
\textbf{Effort of Adoption}\tab
\sysname is a general testing tool for DBMSs.
It takes three steps to apply \sysname to a new DBMS. 
First, we define the mapping rules between parse tree nodes and semantic tree nodes, and use them as input to the parser generation tool.
We define sixty-five mapping rules for PostgreSQL,
seventy-nine for MySQL, and fifty-nine for SQLite. 
% The definition of the mapping rules for a new DBMS
% only takes less than two hours for the leading author of this paper.
Second, we manually add syntax checks beyond the parser and implement semantic analysis for DBMS-specific features.
Third, we modify DBMS to execute multiple query plans and compare the results.
Table~\ref{tab:addoption_effort} lists the human effort of adopting our tool to specific DBMSs.
% As the query optimizers of DBMS are typically different, the effort of modifying DBMS could be DBMS-dependent.
% We only modified 248, 271, and 191 lines of code for SQLite, PostgreSQL, and MySQL, respectively.
% In practice, it took less than one week for the leading author of this paper to adopt \sysname to MySQL.
% \textbf{Compare with NoRec}

\smallskip
\noindent
\textbf{Limitations}\tab
% Our system has several limitations. 
% First, MPE cannot detect logic bugs in non-deterministic queries since it is based on the assumption that the query result is deterministic. That is, all query plans are expected to return the same results. 
First, MPE requires non-empty results to identify logic bugs.
Unfortunately, \sysname does not evaluate the query results during query generation.
We plan to introduce solvers to evaluate the expressions in queries in future work.
% such that the condition of expressions yields true for specific query rows.
Second, MPE may suffer from query plan explosion because the number of possibilities increases dramatically as the number of subqueries or joined tables increases.
Currently, we limit the number of subqueries and joined tables to avoid this issue.
Third, MPE may waste efforts on similar query plans, hurting test efficiency.
We observe that many different queries share some of the same query plans.
% This problem is more pronounced in DBMS with various query optimization, like PostgreSQL.
For the latter two problems, we plan to measure query plan coverage in the DBMS and use it to determine which query plans need to be executed.

\section{Related Work}

\noindent
\textbf{DBMS non-memory bugs detection.}
% DBMS non-memory bug detection relies on test oracles to identify unexpected behavior.
We can classify non-memory bug detection approaches into three categories.

The first approach is based on differential testing which discovers bugs by executing a given input with different DBMS systems.
Slutz proposed RAGE~\cite{slutz1998massive} for finding logic bugs in DBMS by running the same query on different DBMS and comparing their results.
Jinho et al. developed APOLLO \cite{jung2019apollo}, a system to find performance regression bugs by executing the same query on the DBMSs with different versions. 
However, RAGE can only be applied to common features of different DBMSs, and APOLLO can only detect bugs introduced or fixed by newer versions.

Another approach is based on metamorphic testing which identifies bugs by running two queries with equivalent functionality on the same DBMS system.
% Rigger et al. proposed two testing approaches, namely NoREC~\cite{rigger2020detecting} and TLP~\cite{rigger2020finding}.
NoREC~\cite{rigger2020detecting} generates equivalent queries by shifting the conditions in the \code{WHERE} clause to the \code{SELECT} expression. 
% These two queries should have the same set of rows to be evaluated as TRUE.
TLP~\cite{rigger2020finding} partitions a query lacking \code{where} clause into three subqueries whose \code{where} clause are \code{x IS TRUE}, \code{x IS FALSE}, and \code{x IS NULL}.
% Each subquery computes a part of the result, and their combined result is expected to be the same as the original one.
Both of them are limited to the queries that can be converted.

The last approach tries to build the test case along with the corresponding ground truth result.
ADUSA~\cite{khalek2008query} generates all data and the full expected result for a query.
% While this approach could reproduce known bugs and find a new bug in Oracle Database,
However, generating full expected results as ground truth can be expensive which inhibits it from finding more bugs.
To simplify the ground truth generation, Pivoted Query Synthesis(PQS)~\cite{rigger2020testing} only partly validates a query's result.
It synthesizes a query that is expected to fetch a single, randomly-selected row.
By checking whether this row is fetched, PQS can detect logic bugs in the DBMS.
% Similar to NoREC, PQS is also mostly limited to finding bugs in \code{WHERE} clauses.

% We present a novel technique called MPE to identify logic bugs in DBMS. This approach can be interpreted as realizing differential testing by comparing the results of DBMS that implement different query optimization approaches. We believe it can find logic bugs in more SQL features than the previous works because it can better explore query plans produced by different combinations of types of optimizations.

\smallskip
\noindent
\textbf{DBMS test cases generation.}
DBMS requires structural inputs to manipulate data in the database. 
Structural input generation mainly falls into two categories: generate-based approaches and mutation-based approaches. 

The generation-based approach~\cite{rigger2020detecting,rigger2020testing,rigger2020finding,slutz1998massive,sqlsmith,binnig2007qagen,khalek2008query,tang2023detecting} is effective in generating syntax-correct test cases since it typically follows a grammar model that describes the format of the input. 
However, these grammar rules are helpless in improving the semantic correctness of the test case.
SQLsmith~\cite{sqlsmith} is a popular DBMS testing tool that can generate syntax-correct test cases from AST.
Despite it having found over 100 memory bugs in popular DBMSs, SQLsmith achieves quite a low accuracy on semantics which might inhibit it from finding bugs hidden in the deep logic.
QAGen~\cite{binnig2007qagen} proves that generating a completely valid query is NP-complete. 
It improves semantic correctness by combining traditional query processing and symbolic execution.
Previous works also try to improve query generation by generating queries that satisfy certain constraints~\cite{khalek2008query}.
They reduce the query generation into the SAT problem, which is subsequently solved by SAT-solver(e.g., Alloy~\cite{alloy}).
% Unlike previous work that generates queries from grammars, \sysname only uses grammars for mutation so it can take advantage of coverage guidance. 

The mutation-based approach typically incorporates execution feedback to explore the deep logic of tested programs.
The general fuzzers~\cite{stephens2016driller,yun2018qsym,chen2018angora,chen2020savior,gan2020greyone} unaware of the input structure.
They can hardly reach the deep logic of DBMSs even incorporating advanced techniques such as taint analysis or symbolic execution.
Grimoire~\cite{blazytko2019grimoire} utilizes grammar-like combinations to synthesize highly structured inputs without the need for explicit grammar, but most of the generated test cases are still syntax invalid.
Recent works manually provide grammar specifications to guide mutation as they guarantee that the generated queries have correct grammar.
For example, Hardik Bati et al.~\cite{bati2007genetic} propose a genetic approach to mutate SQL by inserting, replacing, or removing grammar with the guidance of execution feedback such as query results, query plans, and traces.
\squirrel~\cite{zhong2020squirrel} relies on a customized parser to translate SQL string into a syntax tree and do the mutation on it to preserve correct grammar.
%To improve semantic correctness, it analyzes the relationship of identifiers in SQL to fill in appropriate valid values.
However, they require expert domain knowledge and nontrivial efforts to construct a precise grammar model.
% \sysname can infer the grammar model automatically by analyzing the grammar file of the targeted DBMS which greatly improves the accuracy of the grammar model and saves a lot of implementation efforts.

\section{Conclusion}
This paper presents \sysname, a mutation-based system for detecting both logic and memory bugs in DBMS.
It proposed two key techniques, i.e., context-sensitive instantiation to generate semantically valid SQL queries during mutation and multi-plan execution that can detect logic bugs in the DBMS execution engine.
We developed a prototype system and applied it to three widely used DBMSs. It successfully identified 50 unique bugs.
The further evaluation shows that \sysname outperforms existing tools.

\def\UrlBreaks{\do\A\do\B\do\C\do\D\do\E\do\F\do\G\do\H\do\I\do\J
\do\K\do\L\do\M\do\N\do\O\do\P\do\Q\do\R\do\S\do\T\do\U\do\V
\do\W\do\X\do\Y\do\Z\do\[\do\\\do\]\do\^\do\_\do\`\do\a\do\b
\do\c\do\d\do\e\do\f\do\g\do\h\do\i\do\j\do\k\do\l\do\m\do\n
\do\o\do\p\do\q\do\r\do\s\do\t\do\u\do\v\do\w\do\x\do\y\do\z
\do\.\do\@\do\\\do\/\do\!\do\_\do\|\do\;\do\>\do\]\do\)\do\,
\do\?\do\'\do+\do\=\do\#}
\bibliographystyle{IEEEtran}
\bibliography{reference}

% \begin{figure*}[th]
% \centerline{\includegraphics[width=0.98\textwidth]{Fig/oracle_compare.png}}
% \caption{Oracle Comparsion. We exclude the results of detected logic bugs in PostgreSQL since none of the systems detected logic bugs.}
% \label{fig:oracle_comparsion}
% \end{figure*}

\newpage
\begin{appendices}
% \onecolumn
\section{PoC Generation Strategies}\label{sec:poc_generation}
MPE can only conduct the PoC on \textbf{modified} DBMSs.
To build PoC on \textbf{unmodified} DBMSs, our goal is to make the buggy query plan to be optimal.
The ways to achieve this goal can be classified into three categories.

First, force a query plan to be chosen by using a hint to give DBMS an explicit optimization instruction. For example, in MySQL, the hint 
 FORCE INDEX forces a table scan to use the specified index if possible. If the buggy query plan scans a table with an index and others do not, this hint could force the plan to be chosen.

Second, eliminate more effective query plans. It can be achieved by turning off the optimization strategies, which is feasible since DBMSs typically provide configuration to switch on/off some optimization strategies. Sometimes, using hints (e.g. IGNORE INDEX) can also eliminate some query plans. Besides, rewriting test cases can achieve this goal as well. For example, by removing the CREATE INDEX statement, query plans that use this index will be eliminated.

Third, modify the cost of query plans by adjusting either \textit{action count} or \textit{action cost}.
A query plan is composed of a series of operations and the cost of an operation can be simplified as \textit{action count} * \textit{action cost}.
\textit{action count} is the amount of data to process/access and \textit{action cost} is the overhead of processing one unit of data.
In some DBMS, e.g. PostgreSQL, the \textit{action cost} is a serial of parameters that can be set by the user.
To change the \textit{action count}, we can modify expressions in the statement, change the DBMS status, or provide hints for SELECT statements.
For example, we found a buggy query plan that sequence scans a table performs worse than another use index search performed better because the SELECT statement had a condition filter "$a<10$".
Suppose the table has 100 rows and only 10 of the rows satisfy this condition, the index search will perform better since it requires less data access.
If we set the condition to "$a<100$" such that all rows satisfy this condition, the sequence scan will perform better because the index search operation requires the same \textit{action count} as the sequence scan operation but extra \textit{action cost} to search the index.
Modifying the data in the query table to make more data satisfy the expression can achieve the same purpose.
Some DBMS allow us to use hints to intervene in the probability that a logical expression is true.
For example, we can use hints "$likely(a<10)$" to tell SQLite that this condition is likely to be true to increase the \textit{action count} of the index search operation.
\end{appendices}

\end{document}